\documentstyle[aps,preprint]{revtex}
\draft
\begin{document}

\title{Dynamical Properties of an Antiferromagnet near the Quantum Critical 
Point: Application to LaCuO$_{2.5}$}

\author{B. Normand and T. M. Rice}

\address{Theoretische Physik, ETH-H\"onggerberg, CH-8093 Z\"urich, 
Switzerland.}

\date{\today}

\maketitle

\begin{abstract}

	For a system of two-chain spin ladders, the ground state for weak 
interladder coupling is the spin-liquid state of the isolated ladder, but is 
an ordered antiferromagnet (AF) for sufficiently large interactions. We
generalize the bond-operator mean-field theory to describe both regimes, and
to focus on the transition between them. In the AF phase near the quantum 
critical point (QCP) we find both spin waves and a low-lying but massive 
amplitude mode which is absent in a conventional AF. The static 
susceptibility has the form $\chi(T) = \chi_0 + a T^2$, with $\chi_0$ small 
for a system near criticality. We consider the dynamical properties to 
examine novel features due to the presence of the amplitude mode, and compute 
the dynamic structure factor. LaCuO$_{2.5}$ is thought to be such an 
unconventional AF, whose ordered phase is located very close to the QCP of 
the transition to the spin liquid. From the N\'eel temperature we deduce 
the interladder coupling, the small ordered moment and the gap in the 
amplitude mode. The dynamical properties unique to near-critical AFs are 
expected to be observable in LaCuO$_{2.5}$. 

\end{abstract}


\section{Introduction}

	The study of spin ladder systems \cite{rdr} has made rapid progress
since their synthesis \cite{rhatb} and identification. \cite{rrgs} Among the
problems investigated both theoretically and experimentally are 
the nature of the ground state, which is dramatically different for ladders
composed of odd and even numbers of chains, the behavior of holes in ladders,
including their superconductivity, \cite{runatmk} and the properties
introduced by doping of spinless impurities. Further novel physical 
phenomena are found in coupled spin ladders, where the magnetic 
structure of even-chain ladders with unfrustrated coupling, is of particular 
interest because the ground state is expected to be a spin liquid at
sufficiently weak coupling, but to order magnetically when the interladder
interactions are strong. This system realizes the 
conditions required to investigate the QCP, 
introduced by Chakravarty {\it et al.} \cite{rchn} in the context of a
nonlinear $\sigma$-model description of the two-dimensional Heisenberg
AF, and discussed extensively and more generally in Refs. 
\onlinecite{rcsy} and \onlinecite{rss}. In the language of these analyses, 
the interladder coupling is a parameter which tunes the system from 
the renormalized classical regime of long-ranged magnetic order, through the 
QCP to the one-dimensional limit where the spins are 
disordered by quantum fluctuations (quantum disordered). 

	Such a structure is represented in three dimensions by the two-chain
ladder compound LaCuO$_{2.5}$. \cite{rht} Initial experimental studies gave
contradictory results on the nature of the ground state, as static
susceptibility measurements \cite{rht} suggested a spin liquid state with a
spin gap, while nuclear magnetic resonance (NMR) \cite{rmkiahkt} and muon spin
resonance ($\mu$SR) \cite{rkoyyakhtn} measurements indicated a transition 
to a magnetically ordered phase. In a brief study of
both electronic and magnetic properties of the material, it was proposed 
\cite{rnr} that the conflicting observations could be reconciled if the system
was located near the QCP of the transition between the two regimes, and 
this scenario was supported by detailed numerical simulations performed by 
Troyer {\it et al.} \cite{rtzu}

	In this work we will analyze the properties of the system on
both sides of the critical point, employing a mean-field approximation 
to the bond-operator technique which is generalized to the magnetically 
ordered regime. From the evolution of the mode structure through
the transition, we may compute both the thermodynamics and the dynamical 
magnetic properties of the system. In the disordered phase, the excitations 
are triply degenerate magnons with a spin gap which vanishes on approach 
to the QCP. In the AF ordered phase, we find that the modes evolve into 
two spin-wave excitations representing rotations of the ordered moment, 
accompanied by an amplitude mode corresponding to fluctuations in its 
magnitude. This latter mode has a gap which grows continuously with the 
moment, so it will be low-lying only near the QCP, and will develop far 
from the transition into a high-lying excitation which is ignored in the 
dynamics of a conventional AF.

	To characterize the ordered system close to the QCP we focus on the 
static susceptibility $\chi(T)$, which vanishes at $T = 0$ on approach to the
transition.  The results for $\chi(T)$ in the mean-field theory agree well 
both qualitatively and quantitatively by highly accurate Quantum Monte Carlo 
studies carried out by Troyer {\it et al.} \cite{rtzu} Both sets of
calculations support the proposal that LaCuO$_{2.5}$ is indeed an AF close to
the QCP. 

	We examine the dynamic structure factor to find features associated
with the presence of an amplitude mode. This mode couples to neutrons in the 
same way as the transverse excitations, and we show that it should be 
observable in inelastic neutron scattering measurements on single crystals. 
Raman light scattering by the magnetic excitations is also discussed, but in 
this case it is difficult to find signals which may be ascribed unambiguously 
to this mode.

	The outline of this paper is as follows. In section II we develop the
bond-operator formalism for application in the ordered magnetic regime, to 
describe the mode structure of the magnetic excitations and the 
zero-temperature phase transition. In section III we discuss briefly the 
statistics of the magnon excitations, in order to solve the mean-field 
equations at finite temperatures and to deduce the static susceptibility on 
both sides of the transition. We consider in section IV the dynamical magnetic
properties of the system in the vicinity of the QCP, and 
calculate the dynamic structure factor for comparison with proposed 
experiments. Section V contains our conclusions and a brief discussion.

\section{Ordered Magnetic Phase}

	We begin by applying the bond-operator technique \cite{rsb,rgrs} in 
the regime where the interladder superexchange coupling is sufficiently strong
to stabilize a magnetically ordered state. We will restrict ourselves to the 
case of simple AF order both along and perpendicular to the ladders. However, 
we shall return in section IV to a recent alternative proposition, and show 
that this gives rise only to notational differences. 

	Following Ref. \onlinecite{rsb}, the spin degrees of freedom 
on each ladder rung are represented by the bond operators
\begin{eqnarray}
| s \rangle & = & s^{\dag} | 0 \rangle = \frac{1}{\sqrt{2}} \left( | \uparrow
\downarrow \rangle  - | \downarrow \uparrow \rangle \right), \label{ebor}
\nonumber \\ | t_x \rangle & = & t_{x}^{\dag} | 0 \rangle = - 
\frac{1}{\sqrt{2}} \left( | \uparrow \uparrow \rangle  - | \downarrow 
\downarrow \rangle \right), \nonumber \\ | t_y \rangle & = & t_{y}^{\dag} 
| 0 \rangle = \frac{i}{\sqrt{2}} \left( | \uparrow \uparrow \rangle  + 
| \downarrow \downarrow \rangle \right), \\ | t_z \rangle & = & t_{z}^{\dag} 
| 0 \rangle = \frac{1}{\sqrt{2}} \left( | \uparrow \downarrow \rangle  + 
| \downarrow \uparrow \rangle \right), \nonumber 
\end{eqnarray}
where the arrows denote the direction of the left and right spins. The 
presence of a quadratic term with negative coefficient in the
transformed Heisenberg Hamiltonian ensures that the singlet operators
condense, so that one may take $\langle s_i \rangle = \overline{s}$ on each
rung. AF ordering along the ladders occurs when, in addition, one of the 
triplet operators has a non-zero expectation value with alternating sign. 
This we take to be the component $t_z$, which may be written as 
\begin{equation}
t_{iz} = (-1)^{i_z} \overline{t} + \hat{t}_{iz} ,
\label{etc}
\end{equation}
where $\hat{t}_z$ contains fluctuations about the average value ${\overline
t}$. The dynamics of these fluctuations may not be neglected, as in the 
treatment of the $s$ degree of freedom, because their dispersion is strong. 
With this choice of basis states, the coexistence of a finite $\overline{t}$ 
with the condensed singlet $\overline{s}$ on a rung corresponds to an AF  
alternation along the ladder of states $ |\uparrow \downarrow \rangle$ and 
$| \downarrow \uparrow \rangle $ of fixed, staggered moment $\overline{t}$, 
admixed with a singlet component of weight $\overline{s} - \overline{t}$. 

	In the disordered or spin-liquid regime the three triplet modes 
remain degenerate and massive, a situation represented schematically in 
Fig. 1(a). The spin gap vanishes as the coupling approaches the QCP, where 
the three modes of the lower branch are spin-wave-like, with a linear 
dispersion about the wave vector of the magnetic ordering (Fig. 1(b)). In 
AF the ordered phase there is a spontaneous breaking of rotational symmetry,
which by Goldstone's theorem is accompanied by massless excitations. These
will be two spin waves, which represent rotations, or transverse oscillations,
of the finite staggered moment. The third mode corresponds to fluctuations in 
the amplitude of the moment (longitudinal) and will acquire a finite gap, as 
shown in Fig. 1(c). In a conventional AF system one observes only the two spin
waves, but close to the QCP the amplitude mode will also be of low energy at 
the zone center. 

	The real-space axes, not to be confused with spin-space labels
introduced above, are chosen with ${\hat {\bf z}}$ along the ladder direction 
and ${\hat {\bf x}}$ and ${\hat {\bf y}}$ the directions of
interladder coupling. Using the reduced unit cell \cite{rtzu} in the $x$- 
and $y$-dimensions, but with doubling of the real-space structure in the 
$z$-direction to accommodate AF ordering along the ladder, the Hamiltonian of 
the three-dimensionally coupled system, 
\begin{eqnarray}
H & = & J \sum_{j} {\bf S}_{l,j} {\bf .S}_{r,j} + \lambda J \sum_{j, m = l,r} 
{\bf S}_{m,j} {\bf .S}_{m,j+{\hat z}} \label{essh} \nonumber \\ 
& & + \lambda^{\prime} J \sum_{j} \left( {\bf S}_{r,j} {\bf .S}_{l,j + 
{\hat x}} + {\bf S}_{r,j} {\bf .S}_{l,j + {\hat y}} \right) , 
\end{eqnarray}
after transformation to bond operators is written as 
\begin{equation}
H = H_0 + H_1 + H_2 + H_3 + H_4 ,
\label{ebosh}
\end{equation}
where the terms have the following origins. 
\begin{eqnarray}
H_0 & = & J \sum_{j,\alpha} \sum_{m = 1,2} \left( - {\textstyle \frac{3}{4}} 
s_{j}^{m \dag} s_{j}^{m} + {\textstyle \frac{1}{4}} t_{j,\alpha}^{m \dag} 
t_{j,\alpha}^{m} \right) \label{ebosh0}
\nonumber \\ & & - \sum_{j,\alpha} \sum_{m = 1,2} \mu_{j,m} \left( 
s_{j}^{m \dag} s_{j}^{m} + t_{j,\alpha}^{m \dag} t_{j,\alpha}^{m} - 1 \right) 
\end{eqnarray}
contains the contribution from the ladder rung interactions, and also the 
constraint which restricts the spin states on each rung to a singlet or
one of three triplets. Here $j$ is a unit cell index and $m$ an index for the
two types of rung in each cell. 
\begin{equation}
H_1 = {\textstyle \frac{1}{2}} \lambda J \sum_{j,\alpha} \sum_{m = 1,2} 
\left( t_{j,\alpha}^{m \dag} t_{j+{\hat z},\alpha}^{m+1} s_{j+{\hat z}}^{m+1
\dag} s_{j}^{m} + t_{j,\alpha}^{m \dag} t_{j+{\hat z},\alpha}^{m+1 \dag} 
s_{j}^{m} s_{j+{\hat z}}^{m+1} 
+ H.c. \right) 
\label{ebosh1}
\end{equation}
and
\begin{equation}
H_2 = {\textstyle \frac{1}{4}} \lambda J \sum_{j,\alpha \ne \beta} \sum_{m =
1,2} \left( t_{j,\alpha}^{m \dag} t_{j+{\hat z},\beta}^{m+1 \dag} t_{j+{\hat 
z},\alpha}^{m+1} t_{j,\beta}^{m} - t_{j,\alpha}^{m \dag} t_{j+{\hat z},
\alpha}^{m+1 \dag} t_{j+{\hat z},\beta}^{m+1} t_{j,\beta}^{m} + H.c. \right) 
\label{ebosh2}
\end{equation}
are the terms quadratic and quartic in $t$ operators corresponding to ladder
leg interactions; $H_2$ may not be discarded in the ordered system because
it will contribute terms of order $\overline{t}^2$ to the dynamics of the
modes $\alpha = x,y$.
\begin{eqnarray}
H_3 & = & - {\textstyle \frac{1}{4}} \lambda^{\prime} J \sum_{j,\alpha} 
\sum_{\eta = \pm {\hat x}, {\hat y}} \sum_{m = 1,2} \left( t_{j,\alpha}^{m 
\dag} t_{j + \eta,\alpha}^{m} s_{j + \eta}^{m \dag} s_{j}^{m} \right. 
\label{ebosh3} 
\nonumber \\ & & \qquad \qquad \qquad + \left. t_{j,\alpha}^{m \dag} t_{j + 
\eta, \alpha}^{m \dag} s_{j + \eta}^{m} s_{j}^{m} + H.c. \right) 
\end{eqnarray}
and 
\begin{eqnarray}
H_4 & = & {\textstyle \frac{1}{8}} \lambda^{\prime} J \sum_{j,\alpha} 
\sum_{\eta = \pm {\hat x}, {\hat y}} \sum_{m = 1,2} \left( t_{j,\alpha}^{m 
\dag} t_{j + \eta, \beta}^{m \dag} t_{j + \eta ,\alpha}^{m} t_{j,\beta}^{m} 
\right. \label{ebosh4} \nonumber \\ & & \qquad \qquad \qquad - \left. 
t_{j,\alpha}^{m \dag} t_{j + \eta, \alpha}^{m \dag} t_{j + \eta, \beta}^{m} 
t_{j,\beta}^{m} + H.c. \right) 
\end{eqnarray}
are the quadratic and quartic contributions from the interladder coupling
terms, which link rungs of the same type $m$.

	In the mean-field approximation the singlet expectation value $\langle
s_{i}^m \rangle = {\overline s}$ and Lagrange multiplier $\mu_{i}^m = \mu$ are
taken to be the same on all rungs $(i,m)$, while the triplet expectation value
$\langle t_{i,z}^m \rangle = - (-1)^m {\overline t}$ alternates. The
constraint term incorporates a reduction of ${\overline s}$ due both to 
the presence of the ${\overline t}$ term and to quadratic fluctuations of all 
three magnon modes. A direct coupling of the longitudinal ($t_z$) fluctuations
to the magnitude of ${\overline s}$ is also relevant, and appears in the 
$t_{z}^{\dag} t_z s^{\dag} s$ terms as 
\begin{eqnarray}
t_{j,z}^{\dag} t_{j+{\hat z},z} s_{j}^{\dag} s_j & = & t_{j,z}^{\dag} 
t_{j+{\hat z},z} \left( 1 - t_{j,z}^{\dag} t_{j,z} \right)^{1/2} \left( 1 - 
t_{j+{\hat z},z}^{\dag} t_{j+{\hat z},z} \right)^{1/2} \label{emt} \nonumber
\\ & \simeq & ({\overline t} + {\hat t}_{j,z}^{\dag}) ( - {\overline t} + 
{\hat t}_{j+{\hat z},z}) \left( {\overline s} - \frac{1}{2 {\overline s}} 
{\hat t}_{j,z}^{\dag} {\hat t}_{j,z} \right) \left( {\overline s} - 
\frac{1}{2 {\overline s}} {\hat t}_{j+{\hat z},z}^{\dag} {\hat t}_{j+{\hat 
z},z} \right) \\ & = & - {\overline s}^2 {\overline t}^2 + {\overline s}^2 
{\hat t}_{j,z}^{\dag} {\hat t}_{j+{\hat z},z} + {\textstyle \frac{1}{2}} 
{\overline t}^2 \left( {\hat t}_{j,z}^{\dag} {\hat t}_{j,z} + {\hat 
t}_{j+{\hat z},z}^{\dag} {\hat t}_{j+{\hat z},z} \right) + O({\hat t}^4) . 
\nonumber
\end{eqnarray}
The first line follows from substitution of the number operators into the
constraint and the second from Eq. (\ref{etc}). In the last line, the first
term is the classical part, the second is a fluctuation term which will appear
in off-diagonal components of the Hamiltonian matrix for all leg and interrung
coupling combinations in $H$ and for all three polarizations ${\hat 
t}_{\alpha}$, and the third is diagonal in the magnon operators so appears as 
a mass term unique to the $t_z$ modes. Note that the mass grows with the
ordered moment ${\overline t}$, as would be expected. 

	The mean-field Hamiltonian in reciprocal space may be cast in the form 
\begin{eqnarray}
H_{\rm m} (\mu, {\overline s}, {\overline t}) & = & N \left( - {\textstyle 
\frac{3}{4}} J {\overline s}^2 + {\textstyle \frac{1}{4}} J {\overline t}^2 
- \mu  {\overline s}^2 - \mu  {\overline t}^2 + \mu - 2 J {\overline s}^2 
{\overline t}^2 (\lambda + \lambda^{\prime}) \right) \label{emfsh} 
\nonumber \\ & & + \sum_{{\bf k} \alpha} \left\{ \sum_{m = 1,2} \left[ 
\Lambda_{\bf k}^{\alpha} t_{{\bf k} \alpha}^{m \dag} t_{{\bf k} \alpha}^{m} 
 + \Delta_{\bf k}^{\alpha} \left( t_{{\bf k} \alpha}^{m \dag} t_{-{\bf k} 
\alpha}^{m \dag} + t_{{\bf k} \alpha}^{m} t_{-{\bf k} \alpha}^{m} \right) 
\right] \right. \\ & & + \left. \left[ \Lambda_{\bf k}^{\prime \alpha} 
t_{{\bf k} \alpha}^{1 \dag} t_{{\bf k} \alpha}^{2} + \Delta_{\bf k}^{\prime
\alpha} \left( t_{{\bf k} \alpha}^{1 \dag} t_{-{\bf k} \alpha}^{2 \dag} + 
t_{{\bf k} \alpha}^{1} t_{-{\bf k} \alpha}^{2} \right) \right] + [ 1 
\leftrightarrow 2] \right\} , \nonumber
\end{eqnarray}
in which $N$ denotes the total number of ladder rungs, the coefficients for 
the transverse modes $\alpha = \sigma \equiv (x,y)$ are
\begin{equation} 
\Lambda_{\bf k}^{\sigma} = {\textstyle \frac{1}{4}} J - \mu - {\textstyle 
\frac{1}{2}} \lambda^{\prime} J ({\overline s}^2 - {\overline t}^2) 
(\cos k_x + \cos k_y) ,
\label{elks}
\end{equation}
\begin{equation} 
\Delta_{\bf k}^{\sigma} = {\textstyle \frac{1}{4}} \lambda^{\prime} J 
({\overline s}^2 + {\overline t}^2) (\cos k_x + \cos k_y) ,
\label{edks}
\end{equation}
\begin{equation} 
\Lambda_{\bf k}^{\prime \sigma} = J ({\overline s}^2 - {\overline t}^2) 
\cos {\textstyle \frac{1}{2}} k_z
\label{eldks}
\end{equation}
and 
\begin{equation} 
\Delta_{\bf k}^{\prime \sigma} = {\textstyle \frac{1}{2}} J ({\overline s}^2 
+ {\overline t}^2) \cos {\textstyle \frac{1}{2}} k_z ,
\label{eddks}
\end{equation}
and for the amplitude modes $\alpha = z$,
\begin{equation} 
\Lambda_{\bf k}^{z} = {\textstyle \frac{1}{4}} J - \mu + 2 J {\overline t}^2 
(\lambda + \lambda^{\prime}) - {\textstyle \frac{1}{2}} 
\lambda^{\prime} J {\overline s}^2 (\cos k_x + \cos k_y) ,
\label{elkz}
\end{equation}
\begin{equation} 
\Delta_{\bf k}^{z} = {\textstyle \frac{1}{4}} \lambda^{\prime} J 
{\overline s}^2 (\cos k_x + \cos k_y) ,
\label{edkz}
\end{equation}
and 
\begin{equation} 
\Lambda_{\bf k}^{\prime z} = 2 \Delta_{\bf k}^{\prime z} = {\textstyle 
\frac{1}{2}} J {\overline s}^2 \cos {\textstyle \frac{1}{2}} k_z .
\label{elddkz}
\end{equation}
The part of $H_{\rm m}$ (\ref{emfsh}) quadratic in the triplet operators
is diagonalized by Bogoliubov transformation, and can be reexpressed in terms
of the appropriate quasiparticle operators $\gamma_{k \alpha}^{\pm}$ as 
\begin{eqnarray}
H_{\rm m} (\mu, {\overline s}, {\overline t}) & = & N \left( - {\textstyle 
\frac{3}{4}} J {\overline s}^2 + {\textstyle \frac{1}{4}} J {\overline t}^2 
- \mu  {\overline s}^2 - \mu  {\overline t}^2 + \mu - 2 J {\overline s}^2 
{\overline t}^2 (\lambda + \lambda^{\prime}) \right) \label{edmfsh} \nonumber 
\\ & & - N \left({\textstyle \frac{1}{4}} J - \mu \right) - {\textstyle
\frac{1}{2}} N \left({\textstyle \frac{1}{4}} J - \mu + 2 J {\overline t}^2 
(\lambda + \lambda^{\prime} \right) \\ & & 
+ \sum_{{\bf k} \alpha} \sum_{\nu = \pm} \omega_{{\bf k} \alpha}^{\nu} 
\left( \gamma_{{\bf k} \alpha}^{\nu \dag} \gamma_{{\bf k} \alpha}^{\nu} + 
{\textstyle \frac{1}{2}} \right) . 
\end{eqnarray}
In calculating the mode frequencies, all terms fourth order in ${\overline s}$ 
and ${\overline t}$ are found to cancel, giving the easily factorized results 
\begin{equation} 
\omega_{{\bf k} \sigma}^{\nu} = \left[ \left({\textstyle \frac{1}{4}} J - 
\mu - 2 \nu J {\overline s}^2 a_{\bf k}^{\nu} \right) \left( {\textstyle 
\frac{1}{4}} J - \mu + 2 \nu J {\overline t}^2 a_{\bf k}^{\nu} \right) 
\right]^{1/2} 
\label{eqsd}
\end{equation}
and 
\begin{equation} 
\omega_{{\bf k} z}^{\nu} = \left[ \left({\textstyle \frac{1}{4}} J - 
\mu + 2 J {\overline t}^2 (\lambda + \lambda^{\prime}) \right) \left( 
{\textstyle \frac{1}{4}} J - \mu + 2 J {\overline t}^2 (\lambda + 
\lambda^{\prime}) - 2 \nu J {\overline s}^2 a_{\bf k}^{\nu} \right)
\right]^{1/2} ,  
\label{eqzd}
\end{equation}
where 
\begin{equation} 
a_{\bf k}^{\pm} = \lambda \cos {\textstyle \frac{1}{2}} k_z \pm {\textstyle 
\frac{1}{2}} \lambda^{\prime} ( \cos k_x + \cos k_y ) .
\label{eak}
\end{equation}
The equation for $\omega_{{\bf k} z}^{\pm}$ represents two branches for the
amplitude mode in the doubled Brillouin zone; folding back of the zone in the
$k_x$ and $k_y$ dimensions returns the four branches in the Brillouin zone of
the real material. The equation for $\omega_{{\bf k} \sigma}^{\pm}$ contains
four doubly-degenerate branches, of which the most interesting is the
lowest-lying, $\omega_{{\bf k} \sigma}^{+}$ with $k_x$ and $k_y$ in the
reduced Brillouin zone. This yields the two rotation modes, and from the 
factorized form (\ref{eqsd}) it is clear that the spin-wave condition is the
same as the vanishing of the spin gap which gave the critical coupling for the
transition from the disordered side, ${\textstyle \frac{1}{4}} J - \mu = 2 J 
{\overline s}^2 (\lambda + \lambda^{\prime})$. 

	That the massless modes described by the bond operators are indeed 
true spin waves can be shown in two ways. First, considering the limit of fully
developed magnetic order, ${\overline s} = {\overline t} = {\textstyle
\frac{1}{\sqrt{2}}}$ and 
\begin{eqnarray}
\omega_{{\bf k} \sigma}^{+} & = & \left({\textstyle \frac{1}{4}} J - \mu 
\right) \left[ 1 - \left( \frac{a_{\bf k}^{+ 2}}{ \lambda + \lambda^{\prime}}
\right)^2 \right]^{1/2} \label{eswl} \nonumber \\ & \simeq & 2 J {\overline
s}^2 \sqrt{ (\lambda + \lambda^{\prime})} \left[ {\textstyle \frac{1}{4}}
\lambda k_{z}^2 + {\textstyle \frac{1}{2}} (k_{x}^2 + k_{y}^2) \right]^{1/2}
\end{eqnarray}
in the limit of small $k$. A textbook derivation \cite{rkqts} of the mode 
spectrum for the Hamiltonian in Eq. (\ref{essh}), for a spins of arbitrary
magnitude $S$ and with four spins per unit cell, returns exactly the contents
of both lines in Eq. (\ref{eswl}), with the condition $S = {\overline s}^2 = 
{\textstyle \frac{1}{2}}$, as required. More generally, in the spin basis of
Eq. (\ref{ebor}) it is straightforward to show that a spin wave, represented
by a staggered, transverse component of $S_x$ on each rung, may be represented
by the bond-operator states $- | t_x \rangle - | s \rangle$, and similarly for
$S_y$, showing its equivalence to a $t_x$ ($t_y$) bond-operator mode in the 
presence of the singlet.  

	The average singlet and triplet occupations may be represented
by the reduced variables 
\begin{equation} 
d_s = \frac{2 J {\overline s}^2}{ {\textstyle \frac{1}{4}} J - \mu}
\;\;\;\;\;\;\;\; d_t \;\; = \;\;  \frac{2 J {\overline t}^2}{ 
{\textstyle \frac{1}{4}} J - \mu} ,
\label{edst}
\end{equation}
in terms of which the mode frequencies (\ref{eqsd}) and (\ref{eqzd}) are given
by  
\begin{equation} 
\omega_{{\bf k} \sigma}^{\nu} = \left({\textstyle \frac{1}{4}} J - 
\mu \right) \sqrt{ ( 1 - \nu d_s a_{k}^{\nu}) ( 1 + \nu d_t a_{k}^{\nu} ) }
\label{eqsdr}
\end{equation}
and 
\begin{equation} 
\omega_{{\bf k} z}^{\nu} = \left({\textstyle \frac{1}{4}} J - 
\mu \right) \sqrt{ \left( 1 + d_t ( \lambda + \lambda^{\prime} ) \right) 
\left( 1 + d_t ( \lambda + \lambda^{\prime} ) - \nu d_s a_{k}^{\nu} \right) },
\label{eqzdr}
\end{equation}
while the spin-wave condition, which is maintained everywhere in the ordered
regime, becomes
\begin{equation} 
d_s = (\lambda + \lambda^{\prime})^{-1}.
\label{ecvds}
\end{equation}
By substitution into the self-consistency equation for the disordered 
solution, \cite{rnr} the critical coupling $\lambda_{c}^{\prime}$ is given 
implicitly by 
\begin{equation} 
\frac{1}{\lambda + \lambda_{c}^{\prime}} = 5 - 3 \sum_{k}^{\prime} 
\left( 1 + \frac{a_k}{\lambda + \lambda_{c}^{\prime}} \right)^{-1/2} n_{\rm m}
(\omega_{\bf k}), 
\label{eztcp}
\end{equation}
where in the reduced unit cell there is only one degenerate magnon branch. In
the remainder of this section we will consider the system at zero temperature,
so that the magnon thermal occupation factors $n_m (\omega_{\bf k}^{\nu})$ 
are unity. The full mean-field equations are 
\begin{eqnarray} 
\langle \frac{\partial H_{\rm m}}{\partial \mu} \rangle & = & 0 = - 
{\overline s}^2 - {\overline t}^2 + 1 \label{emfem} \nonumber \\ & & \;\;
+ {\textstyle \frac{3}{2}} -
{\textstyle \frac{1}{2}} \sum_{{\bf k} \, \nu} ^{\prime} \frac{ 1 + d_t ( 
\lambda + \lambda^{\prime} ) - {\textstyle \frac{1}{2}} \nu d_s a_{\bf 
k}^{\nu} }{ 2 \sqrt{ \left( 1 + d_t ( \lambda + \lambda^{\prime} ) \right) 
\left( 1 + d_t ( \lambda + \lambda^{\prime} ) - \nu d_s a_{k}^{\nu} \right)}}
\\ & & \;\; 
- \sum_{{\bf k} \, \nu} ^{\prime} \frac{ 1 + {\textstyle \frac{1}{2}} \nu d_t 
a_{\bf k}^{\nu} - {\textstyle \frac{1}{2}} \nu d_s a_{\bf k}^{\nu} }{ 2 \sqrt{
( 1 - \nu d_s a_{k}^{\nu}) ( 1 + \nu d_t a_{k}^{\nu})}} \nonumber 
\end{eqnarray}
\begin{eqnarray} 
\langle \frac{\partial H_{\rm m}}{\partial {\overline s}} \rangle & = & 0 = 
- {\textstyle \frac{3}{4}} J - \mu - 2 J {\overline t}^2 (\lambda +
\lambda^{\prime} ) \label{emfes} \nonumber \\ & & \;\;
- {\textstyle \frac{1}{2}} \sum_{{\bf k} \, \nu} ^{\prime} 
\frac{ \nu J a_{\bf k}^{\nu} \left( 1 + d_t ( \lambda + \lambda^{\prime} ) 
\right) }{ 2 \sqrt{ \left( 1 + d_t ( \lambda + \lambda^{\prime} ) \right) 
\left( 1 + d_t ( \lambda + \lambda^{\prime} ) - \nu d_s a_{k}^{\nu} \right)}}
\\ & & \;\; - \sum_{{\bf k} \, \nu} ^{\prime} \frac{ \nu J a_{\bf k}^{\nu} ( 1 
+ \nu d_t a_{\bf k}^{\nu} ) }{ 2 \sqrt{ ( 1 - \nu d_s a_{k}^{\nu}) ( 1 + \nu 
d_t a_{k}^{\nu})}} , \nonumber
\end{eqnarray}
and
\begin{eqnarray} 
\langle \frac{\partial H_{\rm m}}{\partial {\overline t}} \rangle & = & 0 = 
{\textstyle \frac{1}{4}} J - \mu - 2 J {\overline s}^2 (\lambda +
\lambda^{\prime} ) - J(\lambda + \lambda^{\prime} ) \label{emfet} \nonumber 
\\ & & \;\; + {\textstyle \frac{1}{2}} \sum_{{\bf k} \, \nu} ^{\prime} 
\frac{ 2 J (\lambda + \lambda^{\prime}) \left( 1 + d_t ( \lambda + 
\lambda^{\prime} ) - \nu d_s a_{k}^{\nu} \right) }{ 2 \sqrt{ \left( 1 
+ d_t ( \lambda + \lambda^{\prime} ) \right) \left( 1 + d_t ( \lambda + 
\lambda^{\prime} ) - \nu d_s a_{k}^{\nu} \right)}} \\ & & \;\; + \sum_{{\bf k}
\, \nu} ^{\prime} \frac{ \nu J a_{\bf k}^{\nu} ( 1 - \nu d_s 
a_{\bf k}^{\nu} ) }{ 2 \sqrt{ ( 1 - \nu d_s a_{k}^{\nu}) ( 1 + \nu d_t 
a_{k}^{\nu})}} \nonumber ,
\end{eqnarray}
in which the first three terms on the right hand side of each expression
constitute the classical equations, and the remaining terms the quantum
corrections. The notation $\sum ^{\prime}$ denotes ${\textstyle \frac{1}{N}}
\sum$. 

	Unconstrained solution of these equations for $\mu$, ${\overline s}$, 
and ${\overline t}$ at fixed $\lambda$ and $\lambda^{\prime} > 
\lambda_{c}^{\prime}$ gives a set of parameter values which are completely
unrelated to those on the disordered side of the QCP,
meaning a first-order transition. However, such parameters correspond to 
modes which are in general all massive, as the spin-wave condition is not
obeyed, and have no physical meaning in the present problem of an ordered
magnet. This discontinuity between ordered and disordered regimes in an
approach combining classical mean-field terms with quantum corrections has
been encountered previously in similar circumstances. \cite{rss,rs,rtknpc} 
By the Goldstone Theorem, any breaking of a continuous symmetry must be
accompanied by the presence of zero-energy excitations, in this case 
spin waves which are massless in the long-wavelength limit. 
We proceed by enforcing the third mean-field equation (\ref{emfet}) 
classically: this is precisely the spin-wave condition Eq. (\ref{ecvds}), and 
serves both to ensure that the rotation modes are massless at the zone center
and to fix the value of one of the three original variables. 

	The remaining two mean-field equations (\ref{emfem},\ref{emfes}) 
correspond to the pair solved in the disordered phase, now generalized to
include finite ${\overline t}$. These may be combined to yield a single 
equation for $d_t$ as a function of $\delta \lambda^{\prime} = 
\lambda^{\prime} - \lambda_{c}^{\prime}$, which as $d_t \rightarrow 0$ returns
the critical point equation (\ref{eztcp}) for $\lambda_{c}^{\prime}$. 
Solution of this equation gives a weak first-order transition which may be
traced in a linear expansion about $\lambda_{c}^{\prime}$ to the fact that the
small $d_t$ term acts as a cutoff in the ${\bf k}$ summation which provides a 
logarithmic contribution in three dimensions. The linearized equation has the 
form 
\begin{equation} 
d_t ( A + B \ln d_t ) = C \delta \lambda^{\prime} , 
\label{esdtvlp}
\end{equation}
which as shown in Fig. 2, describes reentrant behavior of the onset of 
magnetic order. Numerically, the value ${\overline t}_0$ (Fig. 2) where the 
solution at fixed $\lambda^{\prime}$ becomes double-valued is 0.048. The 
$\lambda^{\prime}$ axis is however grossly expanded, and the reentrance occurs
within the range of values of interladder coupling $ 0.118 < \lambda^{\prime}
< 0.121$. Because the first-order nature of the transition is extremely weak, 
we may use the formalism developed in this section to describe the features 
of the ordered magnetic system both qualitatively an on a quantitative basis. 
Illustration of the zero-temperature solution for the
variation of the ordered moment with the interladder coupling is deferred to
the following section, where it is presented together with the solutions at
finite temperatures. 

\section{Finite-Temperature Solution and Static Susceptibility}

\subsection*{Magnon Statistics}

	To compute the properties of the coupled ladder system at all 
temperatures, it is necessary first to discuss the statistics of the magnon 
excitations, which are contained in the thermal occupation function $n_m 
(\omega_{\bf k}^{\pm})$. \cite{rnr} The essential feature is that despite the
bosonic commutation relations \cite{rsb} of the single magnon operators, 
these do not have conventional bosonic statistics because of the constraint 
(\ref{ebosh0}) on their number, and this has a profound effect on the 
thermodynamics of the system at intermediate and high temperatures. An 
approximate approach which gives a good account of the effects of the
constrained magnon ocupation in the case of the isolated ladder (disordered 
system) was introduced by Tsunetsugu and coworkers in Ref. \onlinecite{rttw}, 
and the reader is referred to this work for a detailed discussion. Here we 
summarize the results of applying this treatment in the ordered magnetic 
regime. 

	The following discussion is simplified by the implicit assumption of
the presence of an anisotropy field stabilizing one particular spatial
direction for the staggered moment, a point to which we shall return in more
detail below when considering the susceptibility. This results in a
small gap in the spin-wave spectrum, preventing the easy axis from reorienting 
in an applied field and allowing for a reduction of the energy of one of the 
modes. 	In an $N$-rung system (${\textstyle \frac{1}{4}} N$ unit cells) with 
both rotation ($\alpha = \sigma \equiv S = \pm 1$) and amplitude ($\alpha = z 
\equiv S = 0$) modes possible at the same wave vector ${\bf k}$, and in the 
presence of a magnetic field $h$, the free energy per site is given by 
\begin{equation} 
\tilde{f} = - \frac{1}{2 \beta} \ln \left\{ 1 + z^z (\beta) + 2 \cosh(\beta h)
z^{\sigma} (\beta) \right\} , 
\label{esfe}
\end{equation} 
where $z^{\alpha} (\beta) = \sum_{\bf k} ^{\prime} {\rm exp} (-\beta 
\omega_{{\bf k}_j \alpha})$, is the partition function for each mode type. 
The magnon thermal occupation functions per mode in zero field are 
\begin{equation} 
\tilde{n}_m (\omega_{{\bf k} \alpha}) = \frac{1}{{\rm e}^{-\beta \omega_{{\bf 
k} \alpha}} + 3} 
\label{emtofd}
\end{equation} 
in the disordered phase where $\omega_{{\bf k} \alpha} = \omega_{{\bf k} z} = 
\omega_{{\bf k} \sigma}$, and in the ordered phase 
\begin{equation} 
\tilde{n}_m (\omega_{{\bf k} \alpha}) = \frac{{\rm e}^{-\beta \omega_{{\bf k} 
\alpha} }}{1 + {\rm e}^{-\beta \omega_{{\bf k} z}} + 2 {\rm e}^{-\beta 
\omega_{{\bf k} \sigma} }} .
\label{emtofoa}
\end{equation} 
That in the latter two equations the occupation function for a mode $z$
($\sigma$) contains a denominator dependent on the energy of a mode $\sigma$ 
($z)$ encodes the effects of the constraint. At high temperatures, each of 
these functions approaches the limiting value of $ {\textstyle \frac{1}{4}}$ 
per mode. The form of the occupation function required in the mean-field 
equations and denoted in Ref. \onlinecite{rnr} as $n_m (\omega_{{\bf k} 
\alpha})$ is that including the zero-point term, {\it i.e.} the analog of the 
bosonic ${\textstyle \frac{1}{2}} \coth \left( {\textstyle \frac{1}{2}} \beta 
\omega_{{\bf k} \alpha} \right)$, and is given by $n_m (\omega_{{\bf k} 
\alpha}) = {\textstyle \frac{1}{2}} + \tilde{n}_m (\omega_{{\bf k} \alpha})$.

\subsection*{Mean-Field Solutions}

	Considering first the approach to the transition from the disordered
side, in Fig. 3 is shown the evolution of the spin gap $\Delta$, the minimum
of the triply degenerate massive magnon dispersion, with interladder coupling
at temperatures between 0 and $0.5J$. The zero-temperature graph is that shown
already in Ref. \onlinecite{rnr}. Solution of the mean-field equations at
finite temperatures is marginally more complicated because the initial system
of two variables may not be reduced to a single one in the same manner. As the
temperature is increased, the disordered phase becomes more robust as might be
expected, and a larger interladder coupling is required to stabilize magnetic
order. The spin gap in the isolated ladder is found also to increase with 
temperature, although this rise is somewhat more linear above $T = 0.1J$ than
would be given by the simplest possible formulation, $\Delta =
\sqrt{\Delta_{0}^2 + T^2}$. 

	In the ordered phase, the characteristic parameter varying with the 
coupling constant is ${\overline t}$, which gives the extent of staggered 
moment formation. This is shown in Fig. 4 for the same temperatures. The
ordered moment rises abruptly at the transition, with no obvious indication 
at any temperature of the reentrant behavior discussed in the previous 
section, and then less steeply thereafter. However, the logarithmic evolution
of the ordered moment given in Eq. (\ref{esdtvlp}) is manifest in a
$\ln$-$\ln$ plot, and precludes the extraction of a power-law mean-field
exponent from data within a range $\Delta \lambda^{\prime} \simeq 0.15$ of the
transition. At high values of the interladder coupling, the ordered moment
falls short of the limiting value ${\overline t} = {\textstyle
\frac{1}{\sqrt{2}}}$ as $\lambda^{\prime} \rightarrow 1$, showing that in the
bond-operator description of the isotropic limit there remains some admixture 
of rung singlets with the ordered spins, a consequence of quantum
fluctuations. 

	Turning now to the physical situation of a fixed interladder coupling
constant at variable temperature, it is clear from Fig. 3 that any system 
will be disordered at high temperature. By solution of a pair of equations 
analogous to the zero-temperature critical-point equation (\ref{eztcp}), this
phase boundary is found in the plane of $T$ and $\lambda^{\prime}$, giving the
N\'eel temperature shown in Fig. 5. Because $T_N$ in this figure is 
measured in units of $J \sim 1400K$, it is evident that the real material 
LaCuO$_{2.5}$, with a N\'eel temperature \cite{rmkiahkt} of 117K, is located 
extremely close to the QCP (in fact at $\lambda^{\prime} = 0.127$). We will 
use for illustration the interladder coupling value $\lambda^{\prime} = 0.13$,
for which the dispersion relations have already been sketched in Fig. 1(c). 
The dispersion curves in the physical Brillouin zone are now shown 
quantitatively in Fig. 6 for this parameter choice, at which the ordered 
moment has the value ${\overline t} = 0.14$ and $T_N = 0.105J$. 

\subsection*{Static Susceptibility}

	The physical quantity most readily measurable which gives direct
information about the magnetic state of a spin system is the static
susceptibility $\chi(T)$, and for this reason it has been the subject of
extensive discussions \cite{rdr,rttw} and measurements in a variety of ladder
systems. \cite{rht,rahtik} The susceptibility per site is  
\begin{equation} 
\chi(T) = - \left. \frac{\partial^2 \tilde{f}}{\partial h^2} \right|_{h=0} ,
\label{essg}
\end{equation} 
which from Eq. (\ref{esfe}) yields in the disordered phase \cite{rttw} 
\begin{equation} 
\chi(T) = \beta \frac{z^{\alpha} (\beta)}{1 + 3 z^{\alpha} (\beta) } .
\label{essd}
\end{equation} 
This is illustrated over a wide temperature range in Fig. 7(a), where it is
clear that the interladder coupling has an effect only at low $T$, while the
remainder of the graph is well described by the numerical results for the
isolated ladder. \cite{rttw} The situation at low temperatures is displayed
in the inset, which shows that as the critical coupling is approached
$\chi(T)$ rises more rapidly with temperature. The nature of
this increase may be studied by examining the function 
\begin{equation} 
g(T) = T^2 \frac{\chi^{\prime} (T)}{\chi(T)},  
\label{essdf}
\end{equation} 
in which $\chi^{\prime}$ denotes the temperature derivative. For the
a system with spin gap \cite{rttw} $\Delta$, $\chi(T) = 
T^{-\alpha} {\rm e}^{- \beta \Delta}$ and $g = \Delta - \alpha T$, while for a
power-law dependence $\chi(T) = a T^{\alpha}$, as might be expected in a
critical regime, $g = \alpha T$. Fig. 7(b) shows this function at low-$T$ 
for several values of the coupling constant $\lambda^{\prime}$, demonstrating
clearly the evolution of the system with interladder coupling from a spin 
liquid with the spin gap $\Delta_0 \simeq 0.5J$ and power $\alpha = {\textstyle
\frac{1}{2}}$ prefactor of the isolated ladder, to a quantum critical phase
in three dimensions, where $\alpha = 2$. These results have a straightforward 
interpretation from the thermal excitations of a dilute gas of (noninteracting)
triplet magnons whose dispersion is quadratic about a spin gap $\Delta$, and
becomes linear as $\Delta \rightarrow 0$. At the QCP the susceptibility is due
to excitations of spin waves, with temperature dependence determined only
by system dimensionality. 

	The calculation of the susceptibility in the ordered phase may proceed
in one of two ways: for an ideal system with no interaction between the spins 
and the real-space or lattice coordinates, the ordered moment simply reorients
in an applied field to be perpendicular, {\it i.e.} a spin-flop transition. 
For a system with an easy-axis anisotropy, the spin waves acquire a mass
proportional to the effective anisotropy field $h_a$, and this allows the 
moment direction to remain stable under application of a parallel field. For
a review of both situations see Ref. \onlinecite{rkhpf}. The susceptibility 
is computed in the zero-field limit, which may always be taken as an external 
field smaller than the intrinsic anisotropy energy, so we restrict our 
attention to the simplest case of the susceptibility contribution from normal 
mode excitations for the case of pinned moment direction in a parallel field.

	In addition to the part due to excitation of normal modes, the 
susceptibility in the ordered phase contains a constant part due to the
presence of the finite moment, and so may be represented as 
\begin{equation} 
\chi(T) = \chi_0 (T) + \chi_{\rm exc} (T) .
\label{essod}
\end{equation} 
$\chi_{\rm exc} (T)$ is computed by analogy with the disordered phase from the
free energy of the modes which may be excited, and is given by 
\begin{equation} 
\chi_{\rm exc}^{\parallel} (T) = \beta \frac{z^{\sigma} (\beta)}{1 + z^{z} 
(\beta) + 2 z^{\sigma} (\beta) } .
\label{esso}
\end{equation} 
Because this part has its leading contributions from the spin-wave modes with
dispersion $\omega_{{\bf k} \sigma}$, it can be expected always to have a
quadratic $T$ dependence at low temperatures. The component due to application
of a perpendicular field is similar, but slightly more involved because the
evolution of the mode $\omega_{{\bf k} z}$ with field is not linear for a
finite gap, a point which will be illustrated below. The numerator contains 
also a term in $z^z (\beta)$, which is much smaller than the spin-wave
contribution when the mass of the $z$ mode is finite, and higher-order terms
which cancel at the transition. The dominant feature of the temperature
dependence remains the $T^2$ part due to the spin-wave contribution, with a
smaller prefactor. On averaging over the crystallite directions in a
polycrystalline sample, $\chi_{\rm exc} (T) = {\textstyle \frac{1}{3}}
\chi_{\rm exc}^{\parallel} (T) + {\textstyle \frac{2}{3}} \chi_{\rm
exc}^{\perp} (T)$. 

	Similarly, the static part in the anisotropy-pinned case may be 
written as $\chi_0 (T) = {\textstyle \frac{2}{3}} \chi_{0}^{\perp}$ for a
polycrystalline sample, as only the transverse part has a finite value and
this will be observed as an average over all crystallite orientations. 
$\chi_{0 \, \perp}$ may be computed in the bond-operator formalism by
considering the additional term in the mean-field Hamiltonian in the presence
of a finite magnetic field ${\bf h}$, 
\begin{equation} 
\sum_i {\bf h.} \left( {\bf S}_{l_i}^1 + {\bf S}_{r_i}^1 + {\bf S}_{l_i}^2 + 
{\bf S}_{r_i}^2 \right) = -i \epsilon_{\alpha \beta \gamma} \sum_i \left( 
t_{i \beta}^{1 \, \dag} t_{i \gamma}^1 + t_{i \beta}^{2 \, \dag} t_{i 
\gamma}^2 \right) . 
\label{ehmft}
\end{equation} 
The evolution of the bond-operator modes of a ladder system with application
of a magnetic field will be presented elsewhere. Here
we state the result that in general a field component $h_{\alpha}$ acts to
create an ordered moment $t_m$ in the operators $t_{\beta}$ and $t_{\gamma}$, 
where $\alpha \ne \beta \ne \gamma$. In the presence of a finite, staggered
$\langle t_z \rangle = {\overline t}$, application of a small field $h_x$ 
stabilizes a finite, staggered ordered moment $\langle - i t_y \rangle \equiv 
{\overline t}_m$. One of the spin-wave modes is unaffected by the field and 
the induced moment and one increases linearly with $h$, while the amplitude 
mode energy rises only quadratically with $h$. We find that ${\overline t}_m
\propto {\overline t} h$, and that the magnetization per site stabilized by 
a small field $h_x$ is ${\overline m}_x = - {\textstyle \frac{1}{2}} 
{\overline t}_m {\overline t}$, giving for the transverse part 
$\chi_{0}^{\perp} = \frac{\partial {\overline m}_x}{\partial h_x} \propto 
{\overline t}^2$. As a function of interladder coupling, this contribution 
appears very similar to the the results in Fig. 5 for ${\overline t}$. 
$\chi_{0}^{\perp}$ vanishes approximately in the manner of a second-order 
transition as the temperature is increased towards $T_N$. When the same
formalism is applied to the case of a longitudinal field, the ordered moment
is $O \left( {\overline t}_{m}^2 \right)$ and it is clear that
$\chi_{0}^{\parallel}$ is vanishing, so that the total $\chi^{\parallel}$ 
has $T^2$ contributions only. 

	In Fig. 8 is shown the full static susceptibility $\chi(T)$
(\ref{essod}) for a variety of values of 
interladder coupling in the ordered phase, $\lambda^{\prime} > 
\lambda_{c}^{\prime}$. The curves begin at the finite value of $\chi_0 (T=0)$
calculated in the preceding paragraphs, and their variation with temperature
is approximately quadratic because $\chi_0 (T)$ is nearly constant at low T,
particularly for values of the interladder coupling not in close proximity to
the critical point. Because the scale of $\chi_0$ is significantly smaller
than that of $\chi_{\rm exc}$, the second-order transition in $\chi_0 (T)$ 
at $T_N$ appears as a minor downward cusp. The results compare very well with
highly accurate Quantum Monte Carlo studies on very large systems by Troyer
{\it et al.} \cite{rtzu}, which show the transition from spin liquid to ordered
magnet occurring very close to $\lambda^{\prime} = 0.12$, and continuous
evolution of the susceptibility from an exponential form in the disordered
phase to a quadratic variation at the critical point with an additive constant
part which grows continuously on moving to the ordered side. Quantitatively, 
the magnitude of the peak susceptibility is some 15\% greater in the
mean-field theory, and that of the constant parts approximately 20\% 
smaller at common values of the coupling in the ordered phase. One may 
fit the measured static susceptibility \cite{rht} to the form $\chi (T) = a 
+ b T^2$ at low temperatures, and quadratic temperature dependence is found 
\cite{rtzu} to give a good account of the data. The intrinsic susceptibility 
$\chi_0$ of the coupled ladder system is very difficult to extract from the
constant $a$, as this requires detailed knowledge of core atomic
susceptibility terms, and we note only the qualitative result that all terms
involved are very small, a further indication of the proximity of the system 
to the QCP. That the mean-field theory is in
such generally good agreement with the numerical results, which for an
unfrustrated spin system and with demonstrably negligible finite-size
corrections can be taken to be essentially exact, is presumed to be primarily
a consequence of the three-dimensionality of the ordered magnetic system. 

\section{Dynamic Magnetic Properties}

	In the preceding section we have deduced that the LaCuO$_{2.5}$ 
system is magnetically ordered but located close to the QCP, as a
result of which the transition temperature and ordered moment are small.
One consequence is that the magnetic modes corresponding to fluctations 
in the amplitude of the ordered moment, which in a conventional magnet are 
high-lying and play no role in determining the low-energy properties of the 
system, should have only a small mass here. In this section
we study the dynamic magnetic properties of such a system to isolate those
features due to the presence of the amplitude mode and predict the ways in
which these may be identified definitively by experiment.

	We focus directly on the dynamic structure factor measured by 
neutron scattering, which may be written following Ref. \onlinecite{rstns} 
in the form 
\begin{eqnarray}
S^{R} ({\bf q},\omega) & = & \frac{1}{2 \pi} \sum_{\alpha \beta} \left( 
\delta_{\alpha \beta} - {\hat q}_{\alpha} {\hat q}_{\beta} \right)
\sum_{i, \langle ij \rangle, \tau, \tau^{\prime}} {\textstyle \frac{1}{4}} 
g_{\tau} g_{\tau^{\prime}} F_{\tau}^{*} ({\bf q}) F_{\tau^{\prime}} ({\bf q})
\label{esfm} \nonumber \\ & & \times
\int_{-\infty}^{\infty} {\rm d} t {\rm e}^{- i \omega t} \langle \exp \left(- 
i {\bf q.r}_{i, \tau} (0) \right) \exp \left( i {\bf q.r}_{i + {\bf r}_{ij}, 
\tau^{\prime}} (t) \right) \rangle \langle S_{i + \tau}^{\alpha} (0) S_{i + 
{\bf r}_{ij} + \tau^{\prime}}^{\beta} (t) \rangle , 
\end{eqnarray}
Here $i$ and $j$ denote unit cells, and ${\bf \tau}$ and ${\bf \tau^{\prime}}$
are vectors specifying the locations of the atoms in each cell.
$F_{\tau} ({\bf q})$ is the magnetic form factor which describes the spatial 
extent of the spin density around the site ${\tau}$, and $g_{\tau}$ is the
Land\'e factor. In this expression the space and spin variables are partially
factorized, and each expectation value can be separated into a constant part, 
corresponding to elastic scattering, and a time-dependent part responsible 
for inelastic processes. 

	We consider first the elastic magnetic component, in order to isolate
those Bragg peaks with the highest intensity about which one may measure the
dynamical structure factor most readily, and also to discuss the nature of the
magnetically ordered state. In Ref. \onlinecite{rnr} it was assumed that the
interladder coupling in LaCuO$_{2.5}$ was AF. This
assumption was made on the basis that ferromagnetic coupling, which relies on
Hund's Rule coupling between orbitals on an atom in the superexchange path, is
generally weaker than AF coupling energy scales, and that 
because the system is one of low symmetry is was likely that the latter would 
have appreciable components on certain paths. In fact this assumption was not
tested in detail, and the tight-binding fit to the Local Density Approximation
(LDA) bandstructure was made using a single Wannier orbital on each Cu 
site, which was taken to be a mixture (due to the low symmetry) of Cu 
3$d_{x^2 - y^2}$- and 3$d_{3z^2 - r^2}$-centred orbitals in the pyramidal 
CuO$_5$ system. The good
agreement obtained with this procedure appeared to justify the assumptions
made, but did not rule out alternative parameter combinations. The results of
the preceding section of this work provide a more reliable means of
estimating the magnitude of the interladder coupling. Recently a quantitative
effort has been made by Mizokawa {\it et al.} \cite{rmpc} to understand the 
superexchange parameters in LaCuO$_{2.5}$. These authors deduce some of the 
parameters of the electronic structure from measurements made by Cu $2p$ 
core-level spectroscopy, and use these in Hartree-Fock calculations. 
They conclude that the magnetically ordered state of lowest 
energy is that where the AF coupled spins in each ladder are ferromagnetically
coupled between the ladders, and that the magnitude of the interladder 
interaction is less than $0.1J$. Superexchange estimates involving competing
exchange paths are difficult, and can involve significant errors 
due to computing small differences between large numbers. Nevertheless
this is the most systematic study performed on the system to date. We note
that the results of the preceding sections are essentially independent of 
the sign of the coupling ratio $\lambda^{\prime}$, which acts only to exchange
the branch indices $\nu = \pm$. 

	The elastic magnetic scattering is described by the static structure
factor, the time-independent part of Eq. (\ref{esfm}), and has finite 
components 
\begin{equation}
S_{\rm s}^{R} ({\bf G}) = N \langle S^z \rangle^2 \left( 1 - ({\bf G.{\hat
z}})_{av}^2 \right) \left[ {\textstyle \frac{1}{2}} g F ({\bf q}) \right]^2 
{\cal F} ({\bf G})
\label{essff}
\end{equation}
only at the reciprocal lattice vectors ${\bf G}$, which are the magnetic 
Bragg peaks. 
Here ${\hat z}$ is the direction of the ordered spin moment in real space,
and is not yet known. In cuprates, the coupling of the spin system to the 
lattice which determines this direction is generally very weak, although
by comparison with the parent phases of tetragonal high-temperature
superconducting materials the spins may be expected to align along either the
rungs or chains of the ladders. $\langle S^z \rangle$ is the magnitude
of the ordered moment, and for the coupling value $\lambda^{\prime} = 0.13$ 
is of magnitude $\sqrt{2} {\overline t} \simeq 0.2$. Thus the square of this 
quantity yields only a 4\% effect, rendering the elastic scattering rather 
weak close to the QCP. 
\begin{equation}
{\cal F} ({\bf G}) = \sum_{{\bf \tau}, {\bf \tau^{\prime}}} (-1)^{(\tau +
\tau^{\prime})} {\rm e}^{i {\bf G.} ( {\bf \tau} - {\bf \tau^{\prime}} ) }
\label{esffg}
\end{equation}
is the structure function, obtained by summing over the sites in the unit
cell, in which the convention used with the site labels ${\tau}$ is such as to
ensure that each $\uparrow$ and $\downarrow$ spins introduce factors of the
opposite sign, corresponding to the spin density distribution around each
site. The appropriate site labeling for one plane of Cu atoms is shown in
Fig. 9(a) for the simple AF structure function, which we denote as Type I AF, 
and in Fig. 9(b) for the case with ferromagnetic coupling between the AF
ladders, which we denote as Type II AF. 
It is straightforward to calculate ${\cal F} ({\bf
G})$, and the results are expressed by quoting the reciprocal lattice vector
${\bf G}$ of the magnetic Brillouin zone as $(h, k, l)$, with $o$ and $e$
denoting odd and even integers respectively. For the Type I configuration 
\begin{equation}
{\cal F} ({\bf G}) = \left\{ \begin{array}{cc} 64 \cos^2 G_x a_x \sin^2 G_y 
a_y & (e,e,o), (o,o,o) \\ 64 \sin^2 G_x a_x \cos^2 G_y a_y & (e,o,o), (o,e,o) 
\\ 0 & l \,\, {\rm even} \end{array} \right. , 
\label{esfafc}
\end{equation}
while for Type II the results are identical with [$x \leftrightarrow y$]; 
the displacement vector components $a_x$ and $a_y$ are shown in Fig. 9. 
These quantities, normalized to unity, are shown in Fig. 10 for the two 
configurations in any reciprocal-space plane of odd $l$ and for values of $h$
and $k$ between 0 and 10. It is evident that there are Bragg peaks of nearly
maximal amplitude, and so the best points in the reciprocal zone around which 
to investigate the dispersion of the dynamical modes would be for example (2, 
0, $o$) and (0, 1, $o$) in Type I, or (3, 0, $o$), (0, 4, $o$) 
and (5, 1, $o$) in Type II. Because the two configurations differ 
significantly in the locations of strong Bragg peaks, it should be possible 
by diffractometry to determine the sign of the interladder superexchange. 

	Turning to the dynamical structure factor as it might be observed (in
a single-crystal sample) around one Bragg peak, the factor of interest is the 
spin part $\int {\rm d} t {\rm e}^{i \omega t} \langle S_{\bf q}^{\alpha} 
(0) S_{-{\bf q}}^{\beta} (t) \rangle$, from which the spatial dependence on 
the crystal structure has been removed. In order to compute this quantity 
in terms of the bond-operator eigenmodes, we take first the rung 
combinations \cite{rgrs} 
\begin{equation}
S_{{\bf q} \alpha}^{\pm} (t) = \sum_i {\rm e}^{i {\bf q.r}_i} \left[ S_{l_i \,
\alpha} (t) \pm S_{r_i \, \alpha} (t) \right] , 
\label{eborc}
\end{equation}
and then combine the variables on the two rungs in the reduced,
AF ordered unit cell to give 
\begin{eqnarray} 
S_{{\bf q} \alpha}^{+ \pm} & = & - i \epsilon_{\alpha \beta \gamma} \sum_{\bf 
k} \left( t_{{\bf k} + {\bf q} \, \beta}^{1 \dag} t_{{\bf k} \, \gamma}^{1} 
\mp t_{{\bf k} + {\bf q} \, \beta}^{2 \dag} t_{{\bf k} \, \gamma}^{2} \right)
\label{eboucc} \nonumber \\ S_{{\bf q} \alpha}^{- \pm} & = & {\overline s}
\left( t_{{\bf q} \, \alpha}^{1} + t_{- {\bf q} \, \alpha}^{1 \dag} \right)
\pm {\overline s} \left( t_{{\bf q} \, \alpha}^{2} + t_{- {\bf q} \, 
\alpha}^{2 \dag} \right) . 
\end{eqnarray}
The final expression for the structure factor 
\begin{equation}
S^R ({\bf q}, \omega) = \sum_{\alpha} \sum_{\mu \nu \eta \rho = \pm} 
\int_{-\infty}^{\infty} {\rm d} t {\rm e}^{i \omega t} \langle S_{{\bf q} 
\alpha}^{\mu \nu} (0) S_{- {\bf q} \alpha}^{\eta \rho} (t) \rangle
\label{edsff}
\end{equation}
is a sum over the three spin indices (for unpolarized neutron scattering) and 
all sign combinations of $S^{\pm \pm}$, and its evaluation is aided by 
extensive cancellation of terms. For scattering studies at low temperatures, 
where very few excited states are occupied and these only singly, we may 
from their bosonic commutation relations \cite{rsb} compute the thermal 
expectation value by summation over bosonic Matsubara frequencies. The two
types of component which emerge from Eq. (\ref{edsff}) are 
\begin{eqnarray} 
S^{R-} ({\bf q}, \omega) & = & \sum_{\alpha} \pi \left( {\overline s}^2 + 
{\overline t}^2 \right) \left( \cosh 2 \theta_{\bf q}^{\alpha} - \sinh 2 
\theta_{\bf q}^{\alpha} \right) \left[ n_{\bf q}^{\alpha} + \Theta (\omega) 
\right] \delta \left( \omega_{\bf q}^{\alpha} - |\omega| \right) 
\label{edsfnmc} \\ S^{R+} ({\bf q}, \omega) & = & \sum_{{\bf k} \, \alpha 
\ne \beta} \left\{ \left( \cosh 2 (\theta_{{\bf k} + {\bf q}}^{\alpha} 
- \theta_{\bf k}^{\beta}) + 1 \right) n_{\bf k}^{\beta} (n_{\bf k}^{\alpha} + 
1) \delta \left( \omega_{{\bf k} + {\bf q}}^{\alpha} - \omega_{\bf k}^{\beta} 
- \omega \right) \right. \label{edsfnms} \nonumber \\ 
& & + \left. {\textstyle \frac{1}{2}} \left( \cosh 2 (\theta_{{\bf k} + {\bf 
q}}^{\alpha} - \theta_{\bf k}^{\beta}) - 1 \right) \left[ n_{\bf k}^{\alpha} 
+ \Theta (\omega) \right] \left[ n_{\bf k}^{\beta} + \Theta (\omega) \right] 
\delta \left( \omega_{{\bf k} + {\bf q}}^{\alpha} + \omega_{\bf k}^{\beta} - 
|\omega| \right) \right\} , 
\end{eqnarray}
in which the hyperbolic trigonometric functions are the coefficients of the
Bogoliubov transformation, \cite{rnr} and the $n_{\bf k}$ are thermal 
occupation functions. Errors due to the constraint on magnon occupations are
neglible at low temperatures, and may in fact be taken into account fully by
a more general formulation of the scattering expression, which gives different
thermal occupation functions $n_{\bf k}$.  
$S^{R-} ({\bf q}, \omega)$ (\ref{edsfnmc}) 
appears with coefficient $\left( {\overline s}^2 + {\overline t}^2 \right)$
for contributions from the $\sigma$ components of the spins, due to the
presence of the ordered moment in $t_z$, and with coefficient ${\overline s}^2$
otherwise. It has the simple interpretation of scattering processes
involving magnon creation or destruction, and will be seen as a broadened
$\delta$-function line along $({\bf q}, \omega_{{\bf q} \alpha})$
corresponding to the magnon dispersion relations. $S^{R+} ({\bf q}, \omega)$
(\ref{edsfnms}) describes magnon-magnon scattering (first line) and pair 
creation or destruction processes (second line), each involving magnons of a 
different type, and so will have $(\sigma, \sigma)$ and $(\sigma, z)$ 
components which could in principle be distinguished by polarized neutron
scattering. 

	The dynamic structure factor $S^{R-} ({\bf q}, \omega)$ for magnon 
creation ($ \omega > 0$) is shown in Fig. 11, and as stated above is found to 
consist of a series of peaks at the dispersion relations of all magnon
branches. A small broadening is inserted by hand in the calculation. In the 
disordered system (Fig. 11(a)) there are four triply degenerate branches in
the physical unit cell, and the lowest of these has a gap at the bottom of its
band ($q = 0$), as required for a spin liquid. In the ordered system
(Fig. 11(b)), this gap has vanished, and branches are split into their
$\sigma$ and $z$ components. In the lowest-lying branch, the intensity of the 
spin-wave peak diverges as $q \rightarrow 0$ and the magnetic Bragg peak is
approached, while the corresponding $z$ mode appears at the energy $\omega = 
0.29J$ at the chosen value of interladder coupling. This energy splitting is
expected to be resolvable, and so the amplitude mode should be detected as a
neighboring peak, or at least as a significant shoulder to the elastic peak,
and to each of the dispersion branches. 
The component $S^{R+} ({\bf q}, \omega)$ for two-magnon scattering
and pair creation has an intensity between one and two orders of magnitude 
lower than that from magnon creation, and so will be barely detectable. 
Because inelastic neutron scattering couples directly to the one-magnon 
process, it is the best technique to observe a clear signature of the 
low-lying amplitude mode.

	We have considered also the expected form of the intensity 
$I(\omega)$ in a Raman light scattering experiment 
on the magnetically ordered but nearly critical system. Unambiguous 
spectroscopic evidence for the amplitude mode, as in the magnon creation 
signal expected in the neutron scattering cross section, would be provided by 
one-magnon processes with a significant weight. Following Fleury and Loudon, 
\cite{rfl} it is clear that in this system neither the magnetic-dipole nor 
electric-dipole first-order interactions detailed by these authors can 
provide appreciable signals. Concluding for completeness with second-order 
light scattering, we have computed the $\langle ({\bf S}_i {\bf .S}_j) 
({\bf S}_i {\bf .S}_j) \rangle$ correlation function which gives the intensity
in a spin-only model. \cite{rsss} We find a ``3$J$'' peak due to scattering
of zone-boundary magnons in the ladder ($\pm k_z$) direction, and that this
peak is broadened in the presence of an amplitude mode split off from the spin
wave band in an AF system close to the QCP. However, mindful of the facts 
that measured spectra are not fully explained by two-magnon scattering 
considerations, and that these are found to be anomalously broad in other 
cuprate materials, we conclude that Raman scattering is unlikely to be a 
suitable probe of the dynamics of the amplitude mode. 

\section{Conclusion}

	We have presented a theoretical description of the three-dimensionally
coupled spin ladder system which is realized in the insulating compound
LaCuO$_{2.5}$. This spin configuration is such that for weak interladder
magnetic coupling the ground state is the spin-liquid phase characteristic 
of the isolated ladder, while for larger coupling values an unfrustrated,
magnetically ordered state is expected. We employ a representation based on
singlet and triplet bond operators on each rung of the ladders to obtain a 
uniform description of both phases within the same framework. 

	Within the mean-field approach to the bond-operator formalism,
solution at zero temperature gives the interladder coupling at the QCP as 
$J^{\prime} = 0.121J$. The transition is found to 
be very weakly first order, and the increase of the ordered moment to be 
logarithmic in the coupling ratio $\lambda^{\prime} - \lambda_{c}^{\prime}$ 
in the vicinity of the transition, with the mean-field exponent recovered at
higher $\lambda^{\prime}$. Finite-temperature solutions are obtained by
incorporating the constraint on the triplet excited states into an effective
magnon statistics, and show the increase with temperature of the disordered
regime and the critical coupling, allowing the deduction of the N\'eel
temperature. 

	The real material has an ordering temperature extremely small on the
scale of the superexchange parameter $J$ within the ladders, implying that it 
is located very close to the QCP on the ordered side. In 
this case the magnon mode corresponding to amplitude fluctuations of the 
ordered moment has only a small gap at the bottom of its band, and will 
contribute to the low-energy dynamic and thermodynamic properties of the
system. The constant part of the static susceptibility is proportional to the 
square of the ordered moment and is small, reconciling measured susceptibility
data with the fact that the system is ordered. The part of the static
susceptibility due to thermal excitation of modes of the system increases
quadratically with temperature for spin waves in three dimensions. The results
of the mean-field theory are supported by comparison with the detailed Quantum
Monte Carlo studies of Troyer {\it et al.} \cite{rtzu}. We have demonstrated 
that the low-lying amplitude mode, which is 
unique to this type of AF, contributes to the dynamical response 
function. Inelastic neutron scattering is an especially suitable probe for
investigating this mode as it may access one-magnon excitation processes, and 
we have presented an explicit calculation of its effects for comparison with 
experiment.

	In closing, for a system so close to the QCP we may
speculate on ways of controlling the tuning parameter represented by the
interladder coupling in order to pass through the transition. As discussed in
the preceding section, the superexchange processes contributing to the
interladder magnetic interaction are not well understood, but the coupling 
is known to be a decreasing function of the bond angle away from 180$^0$. It 
may be possible by application of hydrostatic
pressure, or better uniaxial pressure along the $x$- or $y$-axis of a single
crystal, to cause an alteration of this bond angle significant enough to
measure, at least as a raising or lowering of $T_N$ as deduced from
${\textstyle \frac{1}{T_1}}$ by NMR. As an alternative to physical pressure it
is possible 
\cite{rhpc} also to apply chemical pressure by substituting other trivalent
atoms for La. In LaCuO$_{2.5}$ the bond angle is found to increase on 
substitution of Y for La, while it is decreased in equal amount by
substitution of Nd. We await with interest more detailed results on the
evolution of the magnetic state, which may provide evidence of the system
being moved through the QCP. However, for a random distribution of
substituents there will be a random distribution of local distortions,
from which one may deduce only an average value for the bond angle and 
interladder coupling, so that interpretation within the above framework will
be less transparent. Finally, as we have shown in the preceding sections, the 
changes in many bulk quantities across the quantum critical regime remain 
essentially smooth, and it will be necessary to adopt a criterion based on the
appearance of the staggered moment or the associated massive mode for
unambiguous identification of the transition point. 

\section*{Acknowledgements}

	We are grateful to Z. Hiroi, T.-K. Ng, S. Sachdev, M. Sigrist, M. 
Troyer and M. E. Zhitomirsky for helpful discussions.

\eject

\begin{figure}
\caption{Schematic representation of triplet excitation mode structure
$\omega_{{\bf k} \alpha}$ in LaCuO$_{2.5}$ system. (a) Disordered (spin
liquid) phase at 
low interladder coupling $\lambda^{\prime} < \lambda_{c}^{\prime}$: 
triply-degenerate, massive modes. (b)  Quantum Critical Point
$\lambda^{\prime} = \lambda_{c}^{\prime}$: triply-degenerate spin waves. 
(c) Magnetically Ordered phase at larger interladder couplings 
$\lambda^{\prime} > \lambda_{c}^{\prime}$, with halved Brillouin zone. }
\end{figure}

\begin{figure}
\caption{Graphical representation of weak first-order part in transition of
${\overline t} (\lambda^{\prime})$, from solution of Eq. 
(\protect{\ref{esdtvlp}}). }
\end{figure}

\begin{figure}
\caption{Spin gap as a function of interladder coupling at several values of 
temperature, illustrating monotonic decrease of $\Delta$ with increasing
$\lambda^{\prime}$ towards a vanishing at the point of transition from
disordered to magnetically ordered phase. }
\end{figure}

\begin{figure}
\caption{Magnitude of ordered monent ${\overline t}$ as a function of 
interladder coupling at several values of temperature, illustrating abrupt
increase from the weak first-order transition at $\lambda_{c}^{\prime}$,
followed by more gradual progress towards nearly-full polarization. }
\end{figure}

\begin{figure}
\caption{N\'eel Temperature $T_N$ as a function of interladder coupling. For
LaCuO$_{2.5}$, the measured value $T_N$ = 117K $<
0.1J$ \protect{\cite{rmkiahkt}} indicates very close proximity to the QCP. }
\end{figure}

\begin{figure}
\caption{(a) Static susceptibility $\chi(T)$ for several values of interladder
coupling in the disordered phase. Deviations from the behavior of the isolated
ladder are observed only at temperatures $T < \lambda^{\prime} J$ below the
scale of the coupling energy. This regime is expanded in the inset. (b) The
function $f(T)$ (\protect{\ref{essdf}}) for four values of $\lambda^{\prime}$
spanning the disordered phase, illustrating the change from exponentially
activated to quadratic susceptibility with decreasing spin gap. }
\end{figure}

\begin{figure}
\caption{Magnon dispersion relations in physical Brillouin zone along the
lines X$\Gamma$ and $\Gamma$Z for magnetically ordered system
$\lambda^{\prime} = 0.13$. All branches $\sigma$ are doubly degenerate, and the
lower $\omega_{\bf k}^{z +}$ line is the amplitude mode with a small mass
(0.29$J$ here) near the QCP. }
\end{figure}

\begin{figure}
\caption{Static susceptibility $\chi(T)$ for several values of interladder
coupling in the ordered phase. The constant part $\chi_0$ rises, at first
rapidly, with $\lambda^{\prime}$ away from $\lambda_{c}^{\prime}$. Small
downward cusps appear at $T = T_N$, marking the second-order vanishing of
$\chi_0$. Isolated-ladder character is recovered at high temperature. }
\end{figure}

\begin{figure}
\caption{Unit cell structure and labeling for one plane of atoms in
magnetically ordered material, for computation of structure function ${\cal F}
({\bf G})$ (\protect{\ref{esffg}}). Intraladder coupling (perpendicular to
plane of diagram) is always AF, whence unit-cell doubling. 
(a) Type I AF configuration. (b) Type II AF configuration. }
\end{figure}

\begin{figure}
\caption{Structure function ${\cal F} ({\bf G})$ at reciprocal lattice points
($h, k, l$=odd), for $0 \le h, k \le 10$. (a) Type I AF. (b) Type II AF. The 
full static structure factor requires also multiplication by the square of
the ordered moment, and so will be rather small for a system close to the 
QCP. }
\end{figure}

\begin{figure}
\caption{Dynamic structure factor $S^R (q_z, \omega)$ expected from
inelastic neutron scattering, illustrating contributions from magnon creation
processes, which appear as a broadened $\delta$-function at the dispersion
branches $\omega_{\alpha} (q_z)$ (a) Disordered system, $\lambda^{\prime} = 
0.05$: all four branches in physical Brillouin zone are triply degenerate, and
there is a gap in excitation energy $\omega$ to the first mode at $q = 0$. 
(b) Ordered system close to the critical point, $\lambda^{\prime} = 0.13$: 
branches are split into $\sigma$ and (massive) $z$ modes. For
the lowest branch, the spin-wave scattering amplitude diverges as $q^2$ as $q
\rightarrow 0$, forming the elastic peak. The signature of the massive but
low-lying amplitude mode is the peak dispersing from $\omega = 0.29J$ at $q =
0$. }
\end{figure}

\newpage

\addtolength{\oddsidemargin}{-1.7cm}
\addtolength{\textwidth}{2.6cm}
\addtolength{\topmargin}{-3.0cm}
\addtolength{\textheight}{7.0cm}
\pagestyle{empty}
\input psfig

\begin{figure}[hp]
\mbox{\psfig{figure=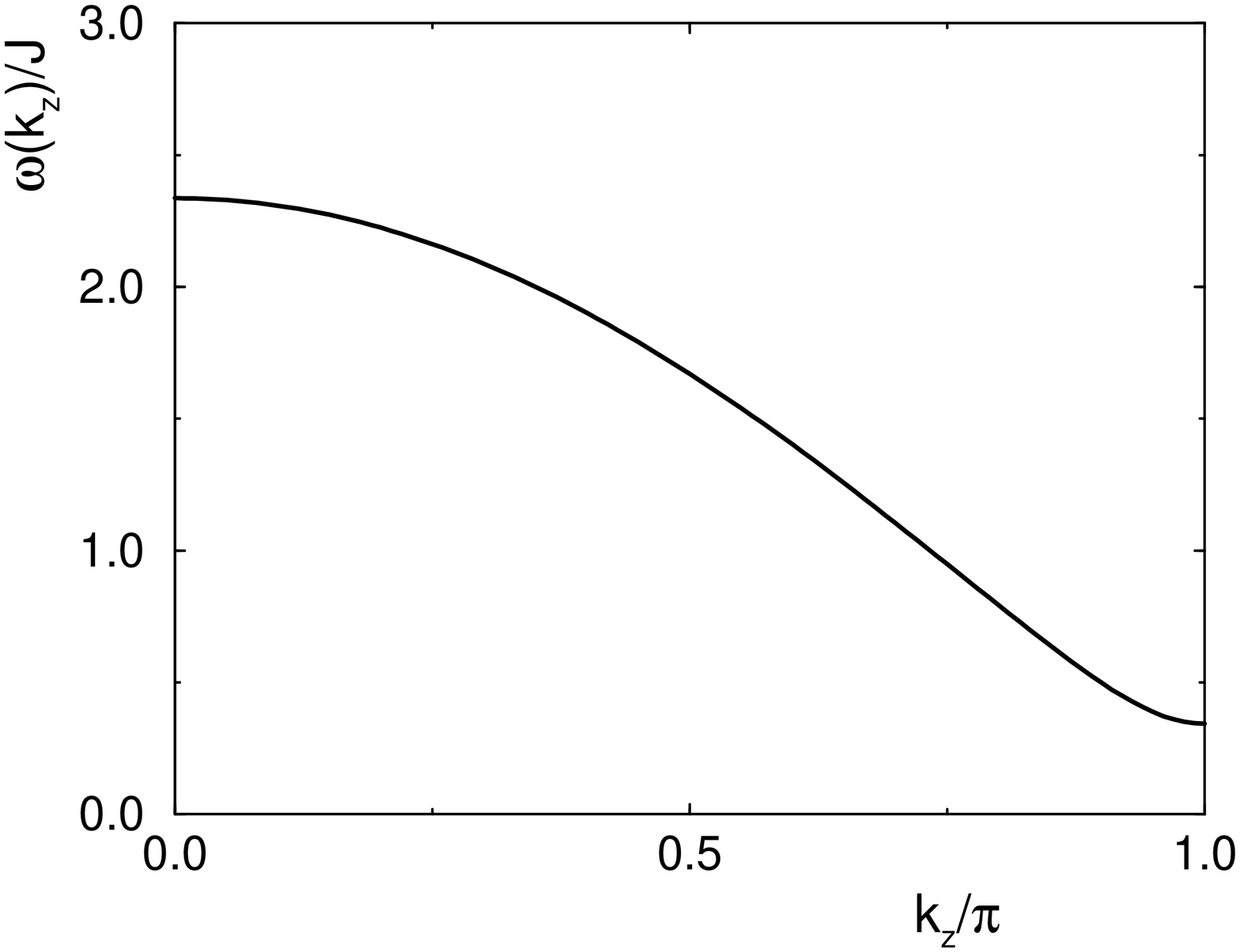,height=6cm,angle=270}}
\mbox{\psfig{figure=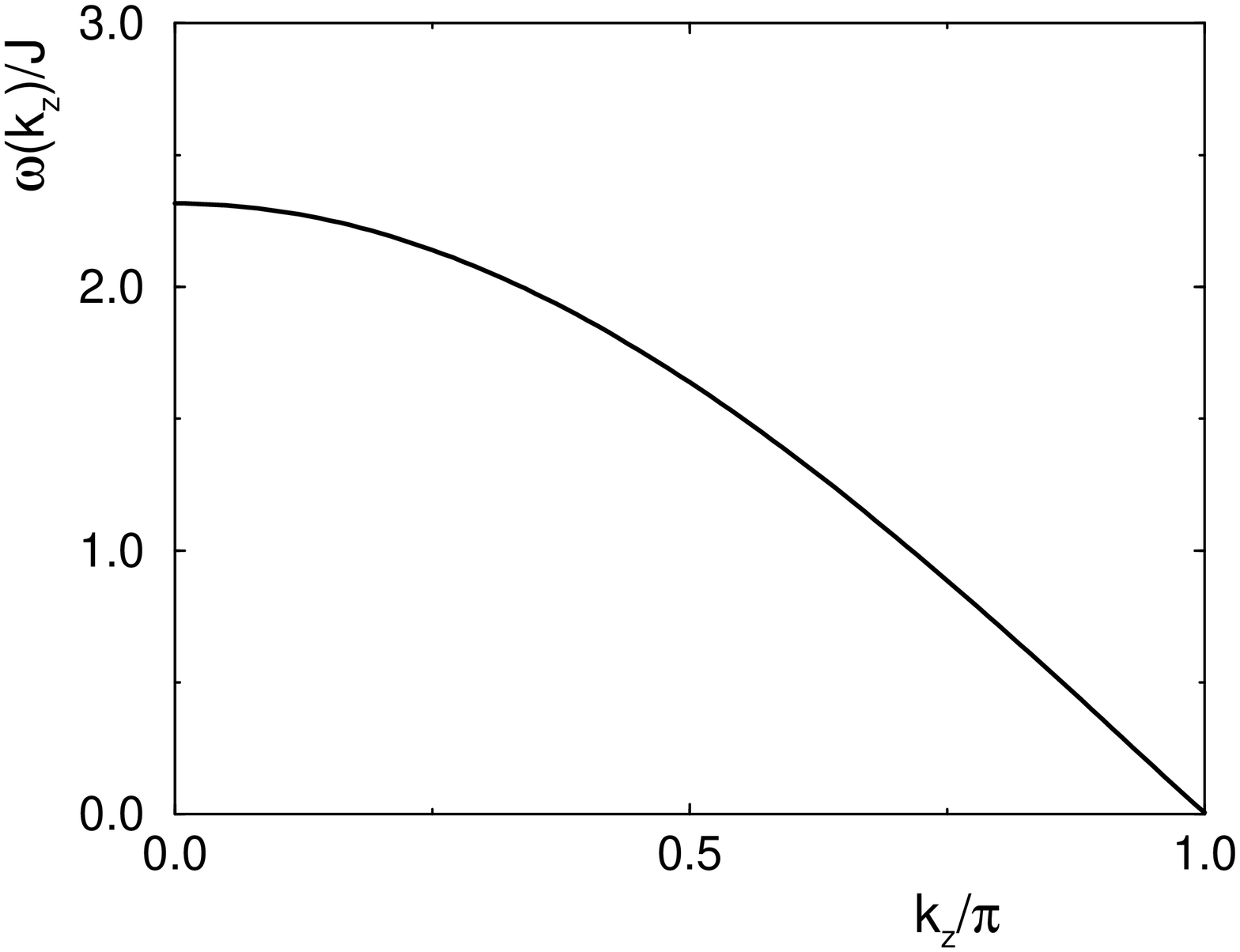,height=6cm,angle=270}}
\end{figure}
\vskip 0.5cm
\centerline{\huge 1(a) \qquad\qquad\qquad\qquad 1(b)}
\vskip 0.5cm

\begin{figure}[hp]
\mbox{\psfig{figure=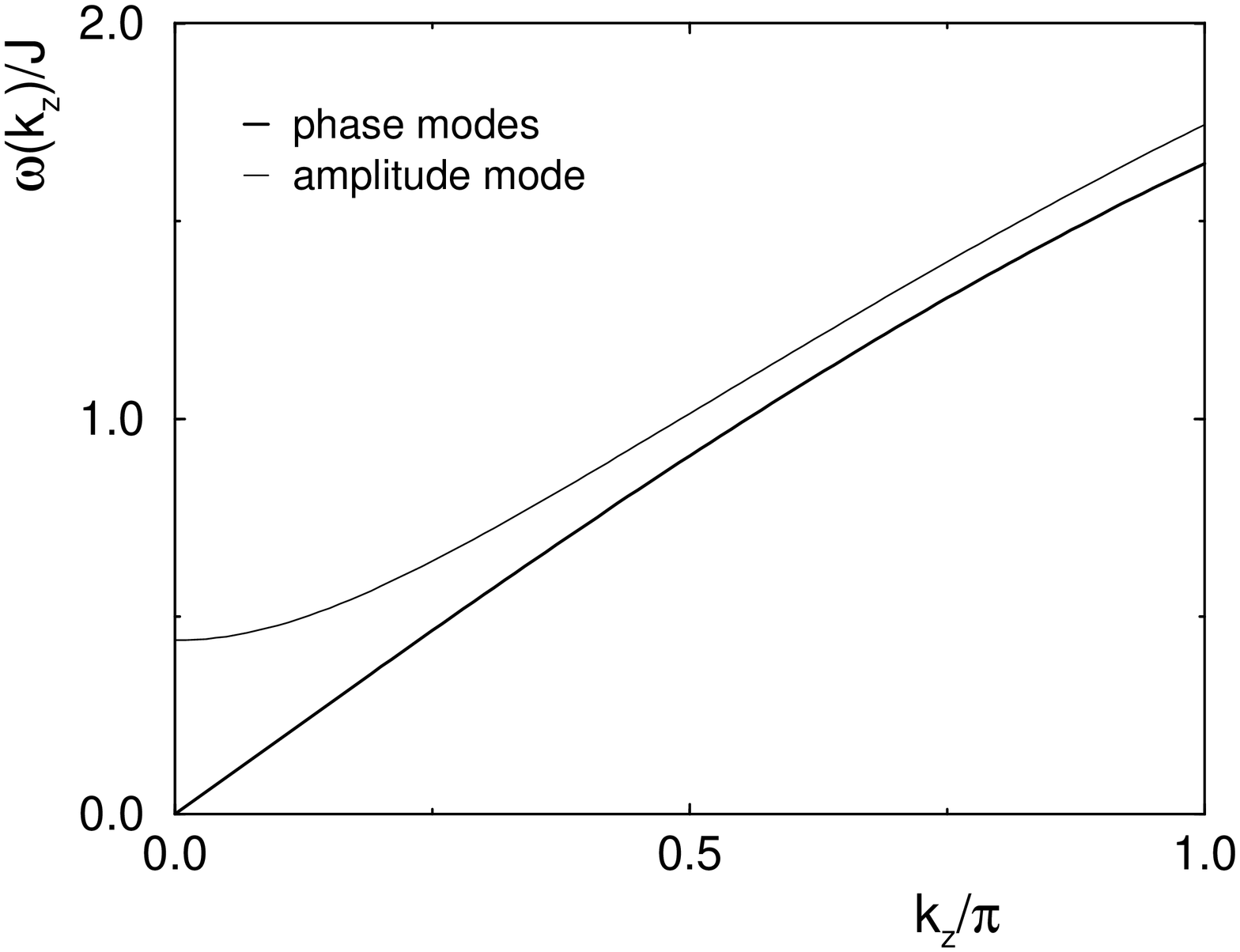,height=6cm,angle=270}}
\mbox{\psfig{figure=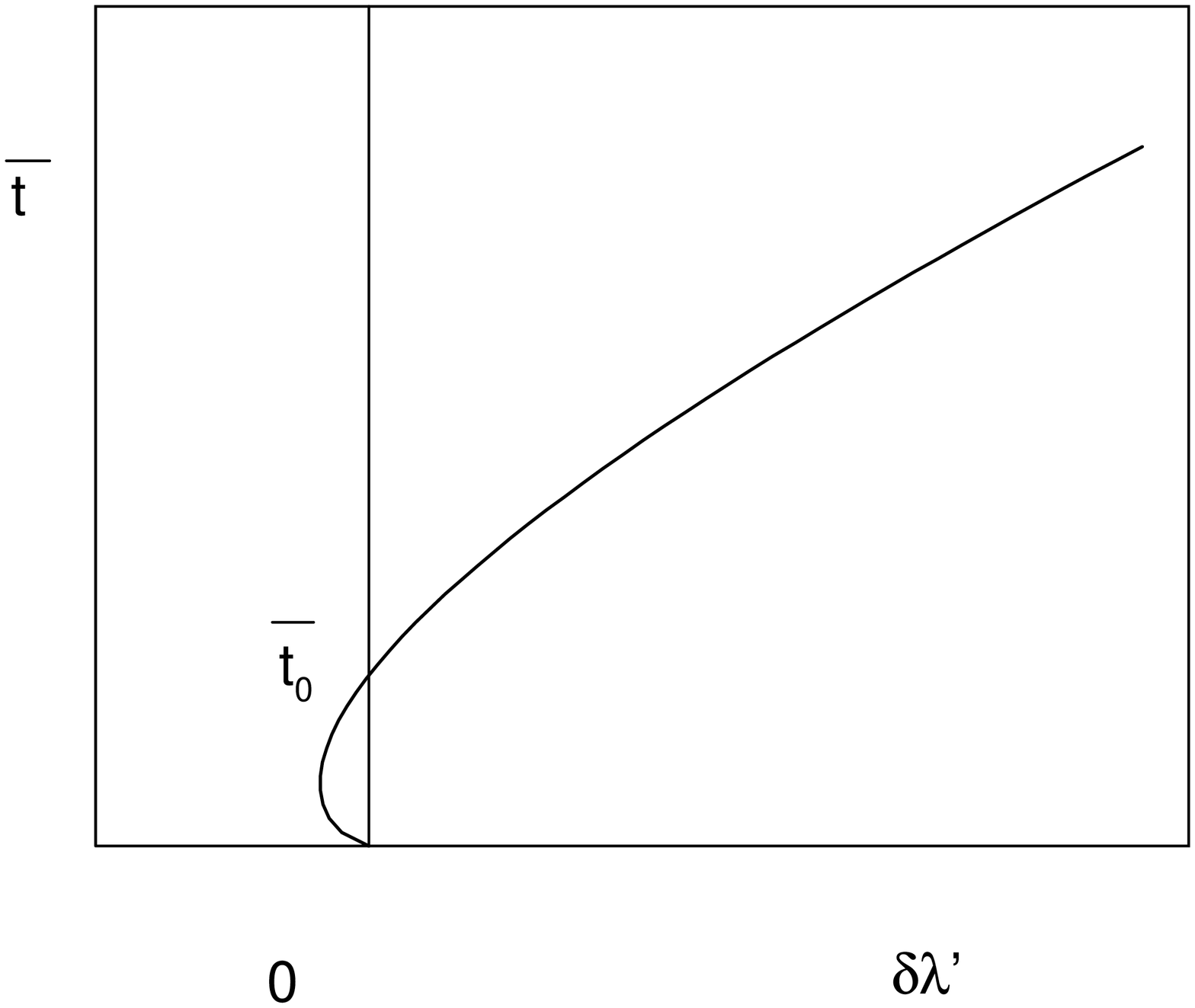,height=6cm,angle=270}}
\end{figure}
\vskip 0.5cm
\centerline{\huge 1(c) \qquad\qquad\qquad\qquad 2}

\vskip 1cm
\centerline{\huge Figs. 1 \& 2, B. Normand and T. M. Rice }

\newpage

\begin{figure}[hp]
\centerline{\psfig{figure=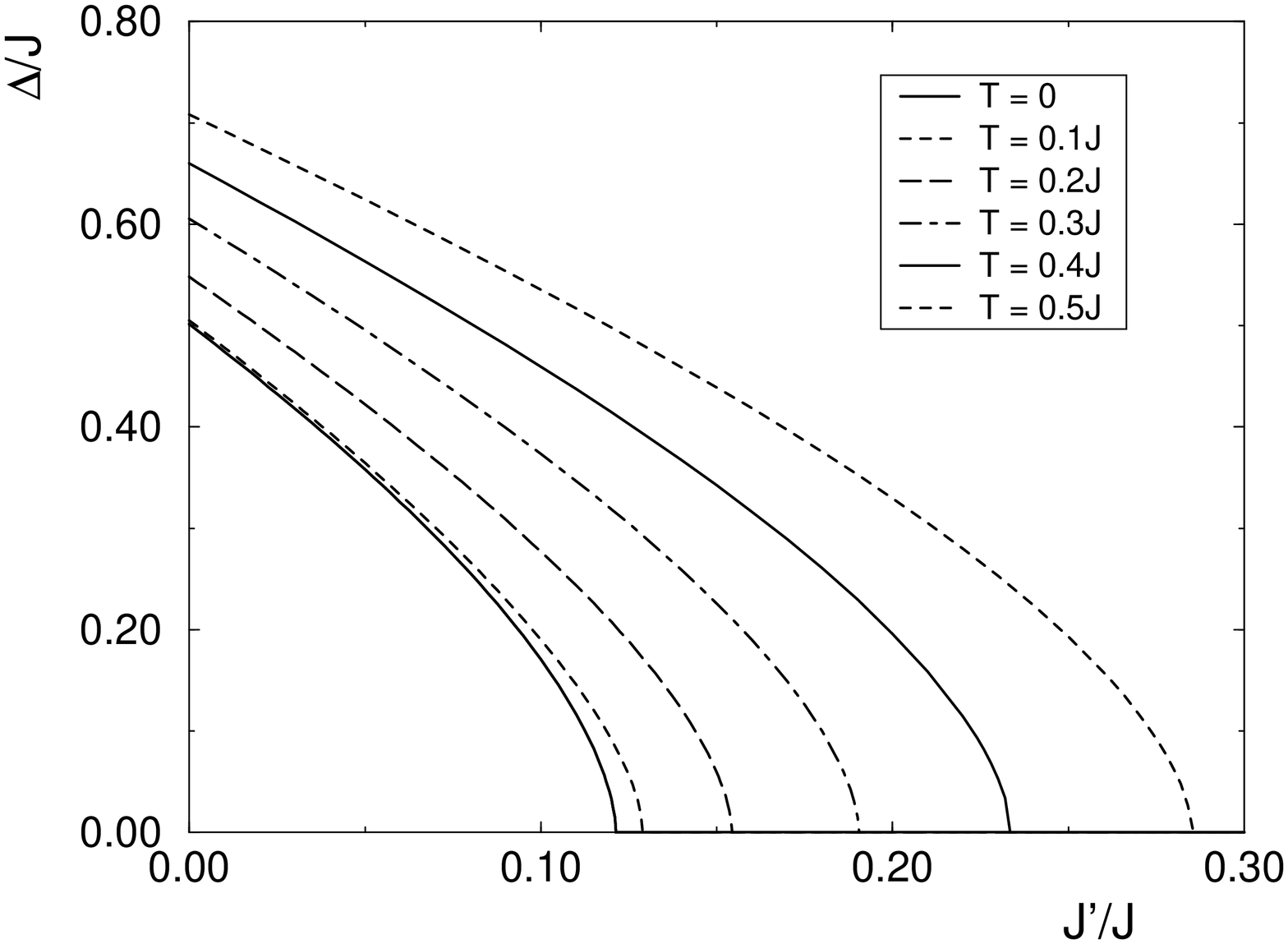,height=9.5cm,angle=270}}
\end{figure}

\centerline{\huge Fig. 3, B. Normand and T. M. Rice }

\begin{figure}[hp]
\centerline{\psfig{figure=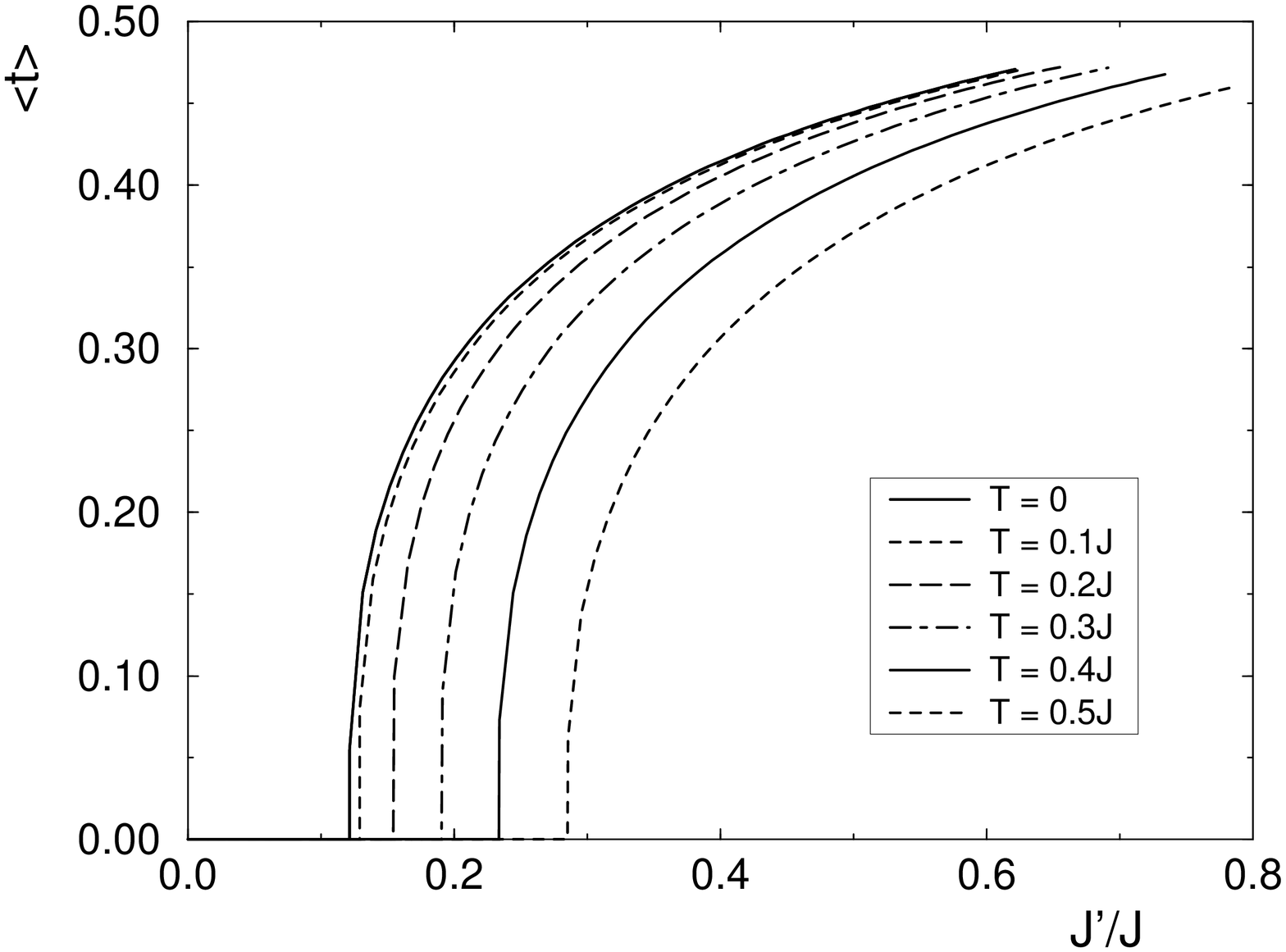,height=9.5cm,angle=270}}
\end{figure}

\centerline{\huge Fig. 4, B. Normand and T. M. Rice }

\newpage

\begin{figure}[hp]
\centerline{\psfig{figure=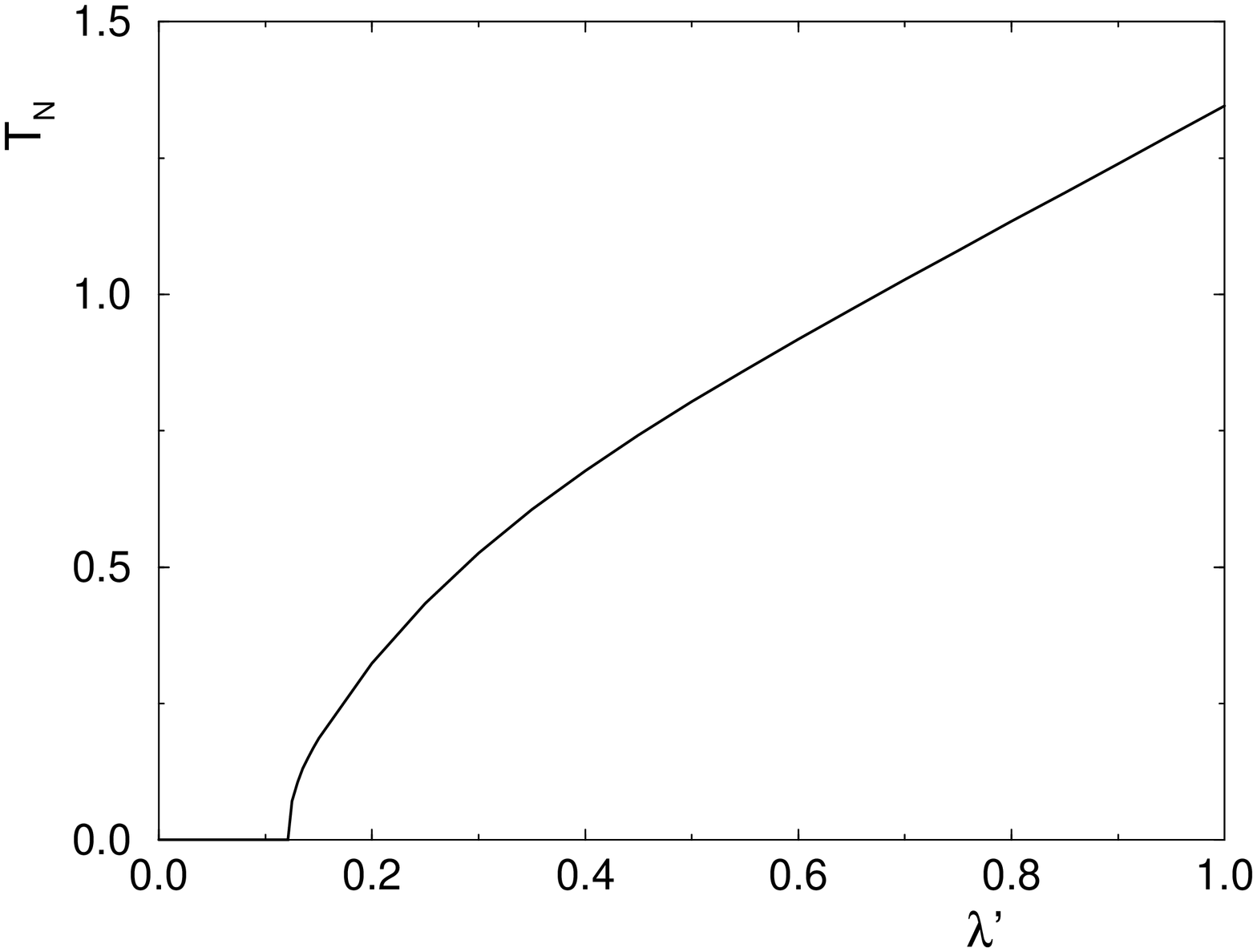,height=9.5cm,angle=270}}
\end{figure}

\centerline{\huge Fig. 5, B. Normand and T. M. Rice }

\begin{figure}[hp]
\centerline{\psfig{figure=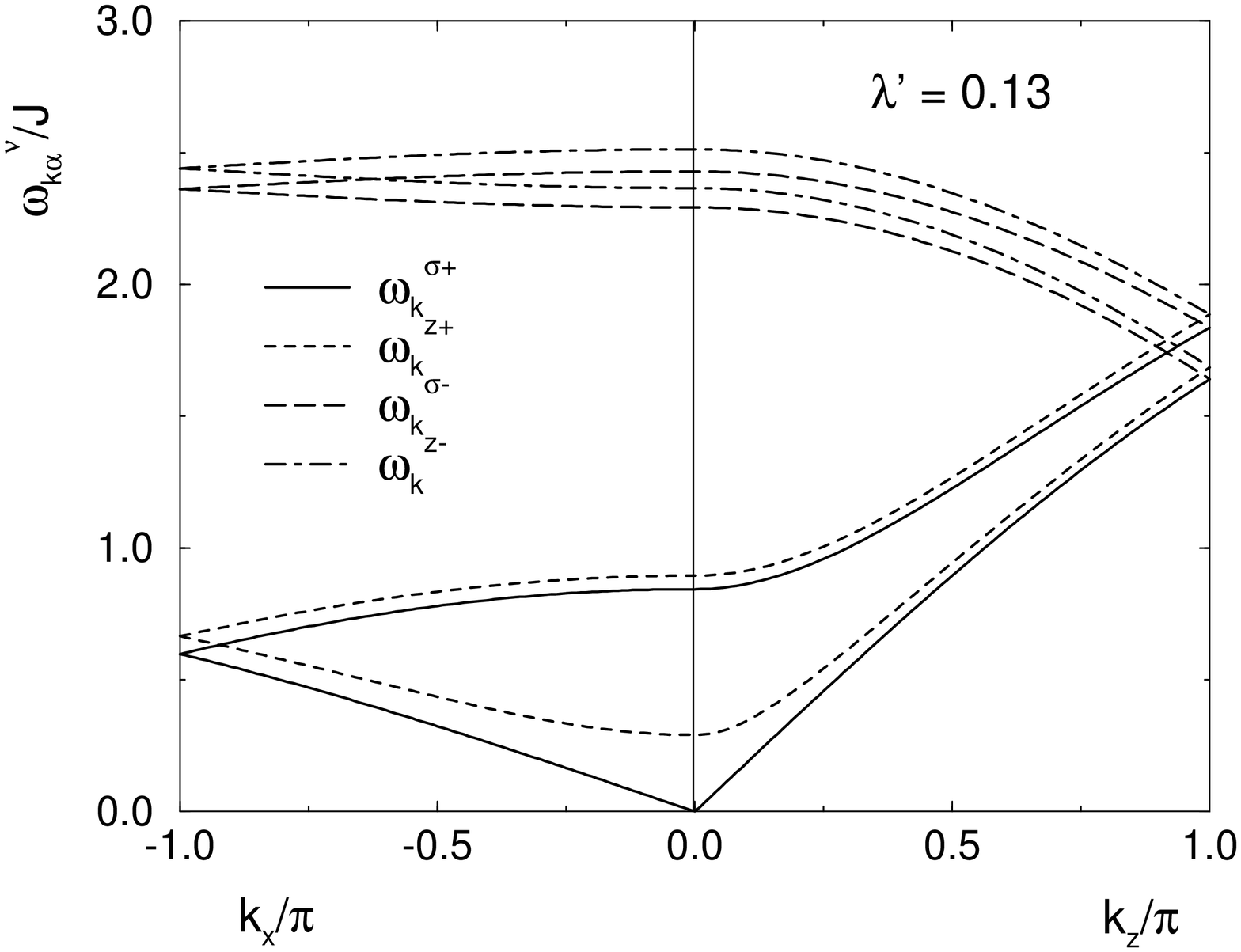,height=9.5cm,angle=270}}
\end{figure}

\centerline{\huge Fig. 6, B. Normand and T. M. Rice }

\newpage

\begin{figure}[hp]
\centerline{\psfig{figure=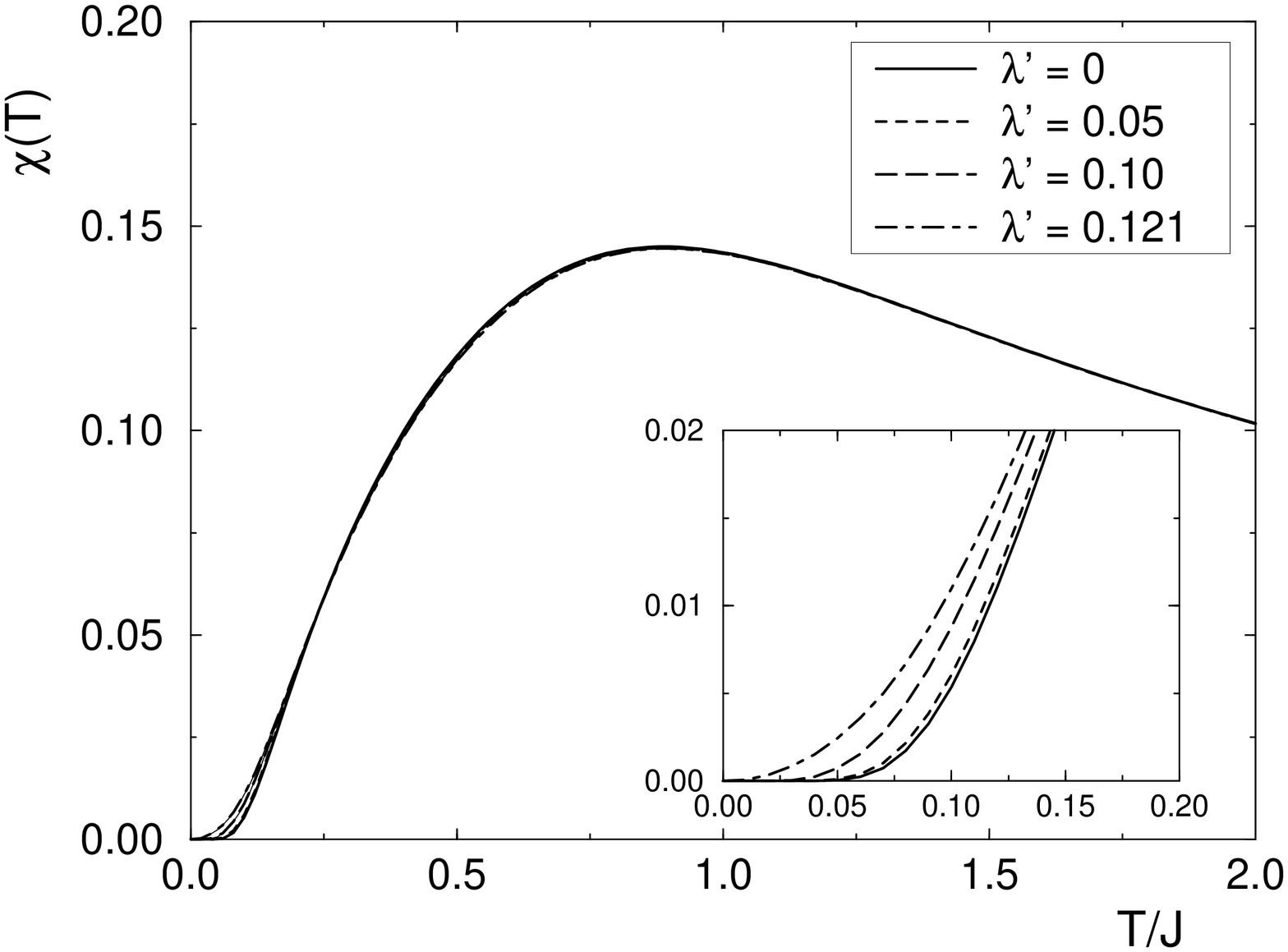,height=9.5cm,angle=270}}
\end{figure}

\centerline{\huge (a)} 

\begin{figure}[hp]
\centerline{\psfig{figure=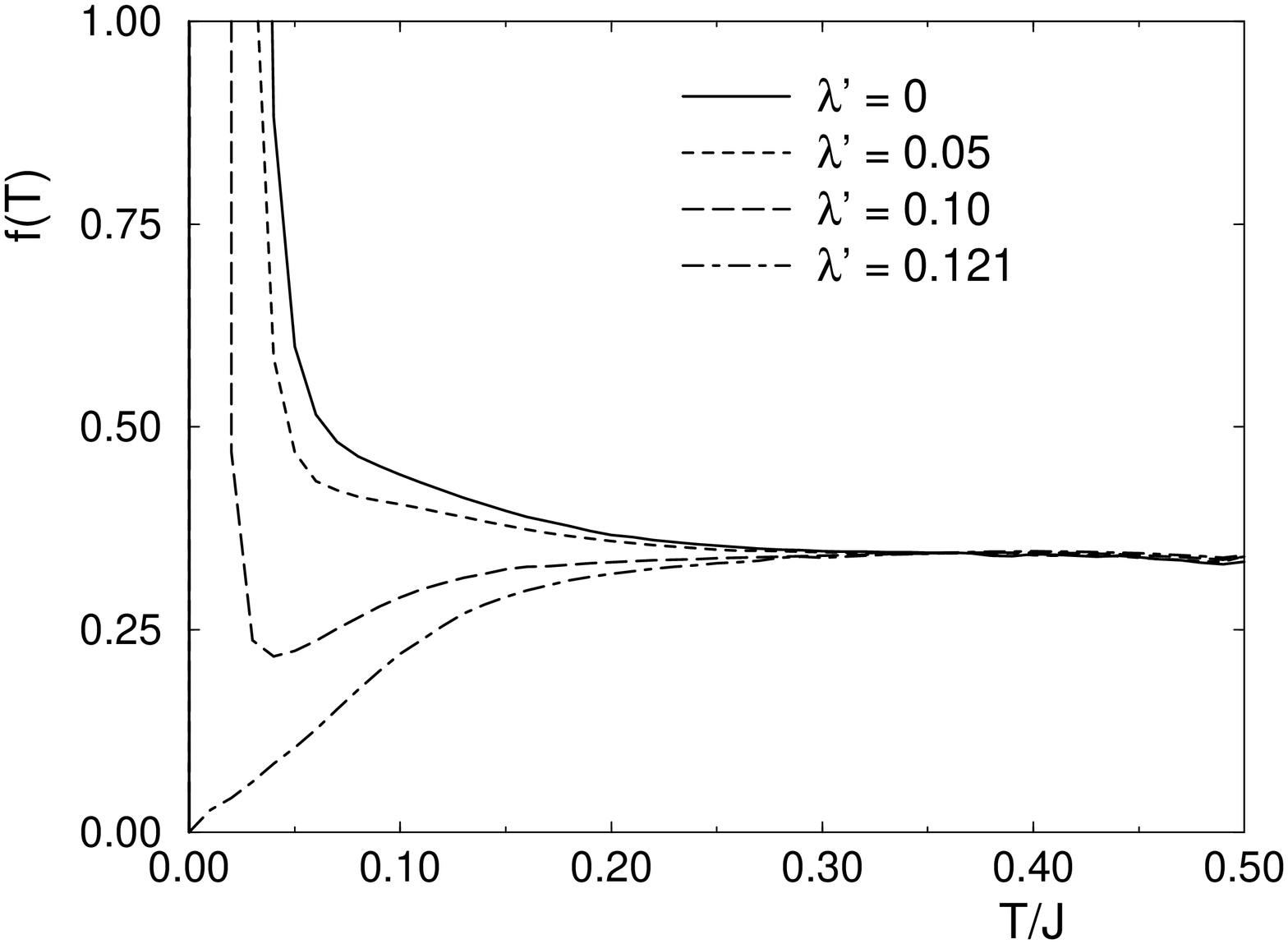,height=9.5cm,angle=270}}
\end{figure}

\centerline{\huge (b)} 
\vskip 0.5cm
\centerline{\huge Fig. 7, B. Normand and T. M. Rice }

\newpage

\begin{figure}[hp]
\centerline{\psfig{figure=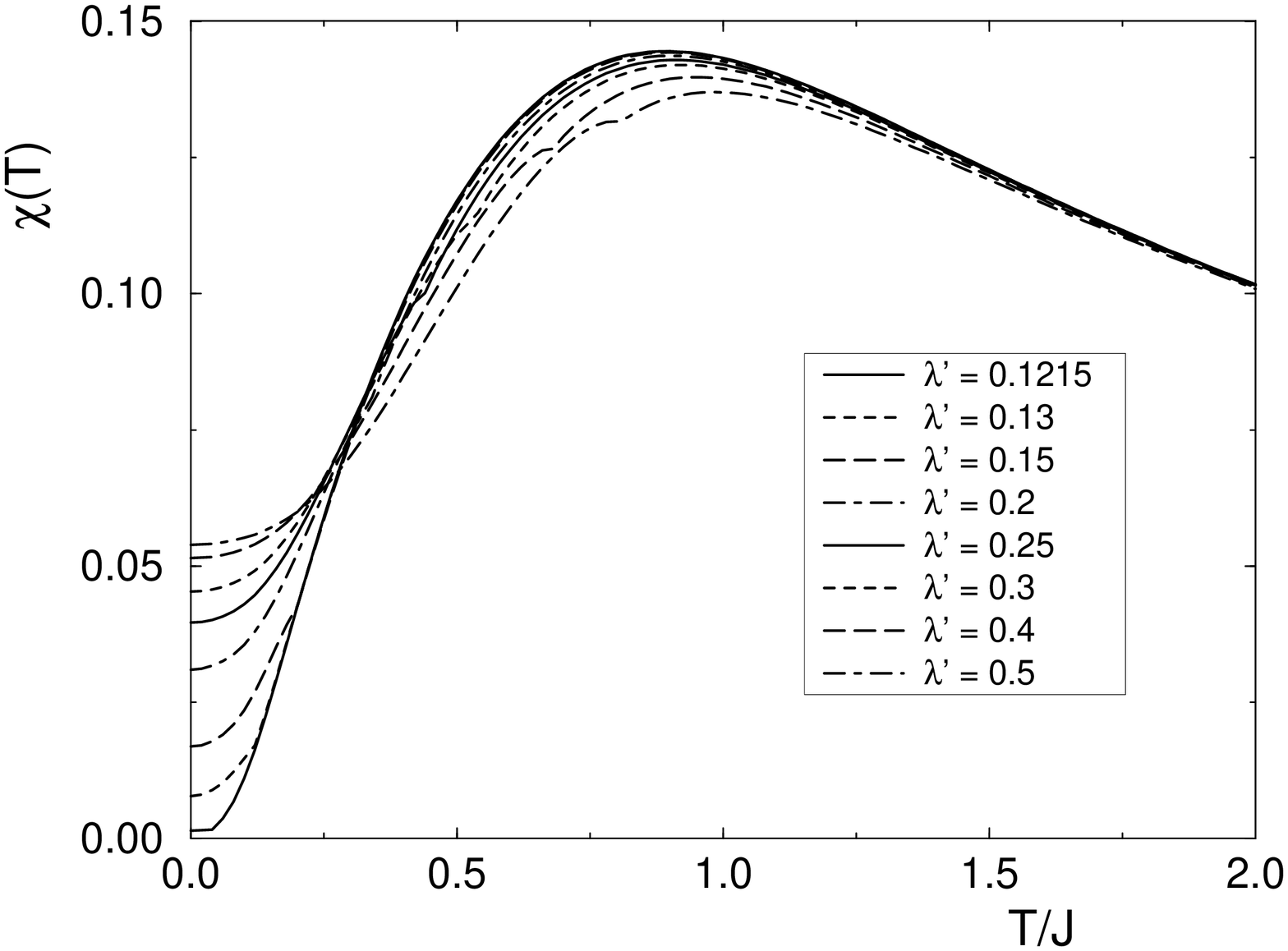,height=9.5cm,angle=270}}
\end{figure}

\centerline{\huge Fig. 8, B. Normand and T. M. Rice }

\vskip 1cm

\begin{figure}[hp]
\centerline{\psfig{figure=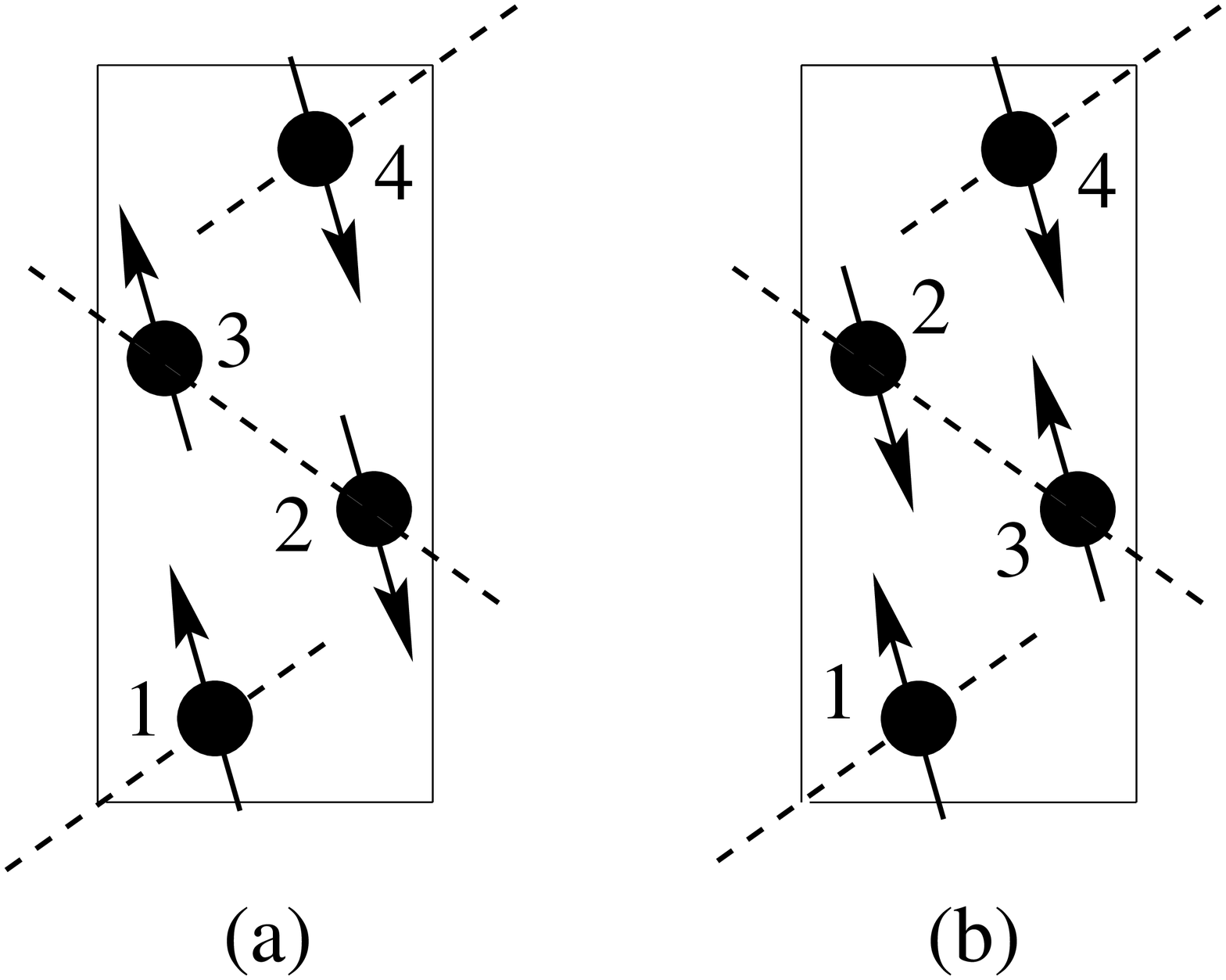,height=9.5cm,angle=0}}
\end{figure}

\centerline{\huge Fig. 9, B. Normand and T. M. Rice }

\newpage

\begin{figure}[hp]
\centerline{\psfig{figure=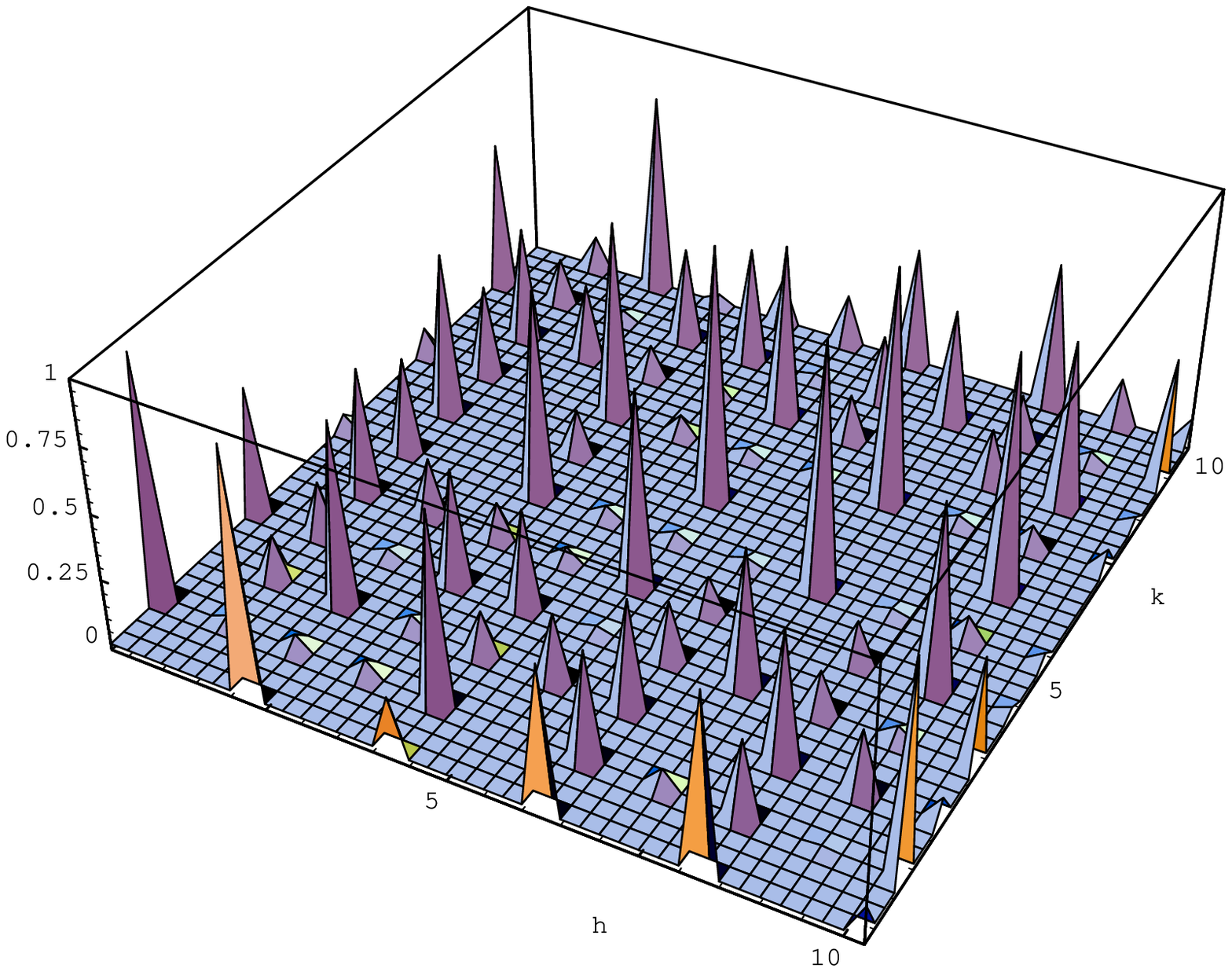,height=9.5cm,angle=0}}
\end{figure}

\centerline{\huge (a)} 

\begin{figure}[hp]
\centerline{\psfig{figure=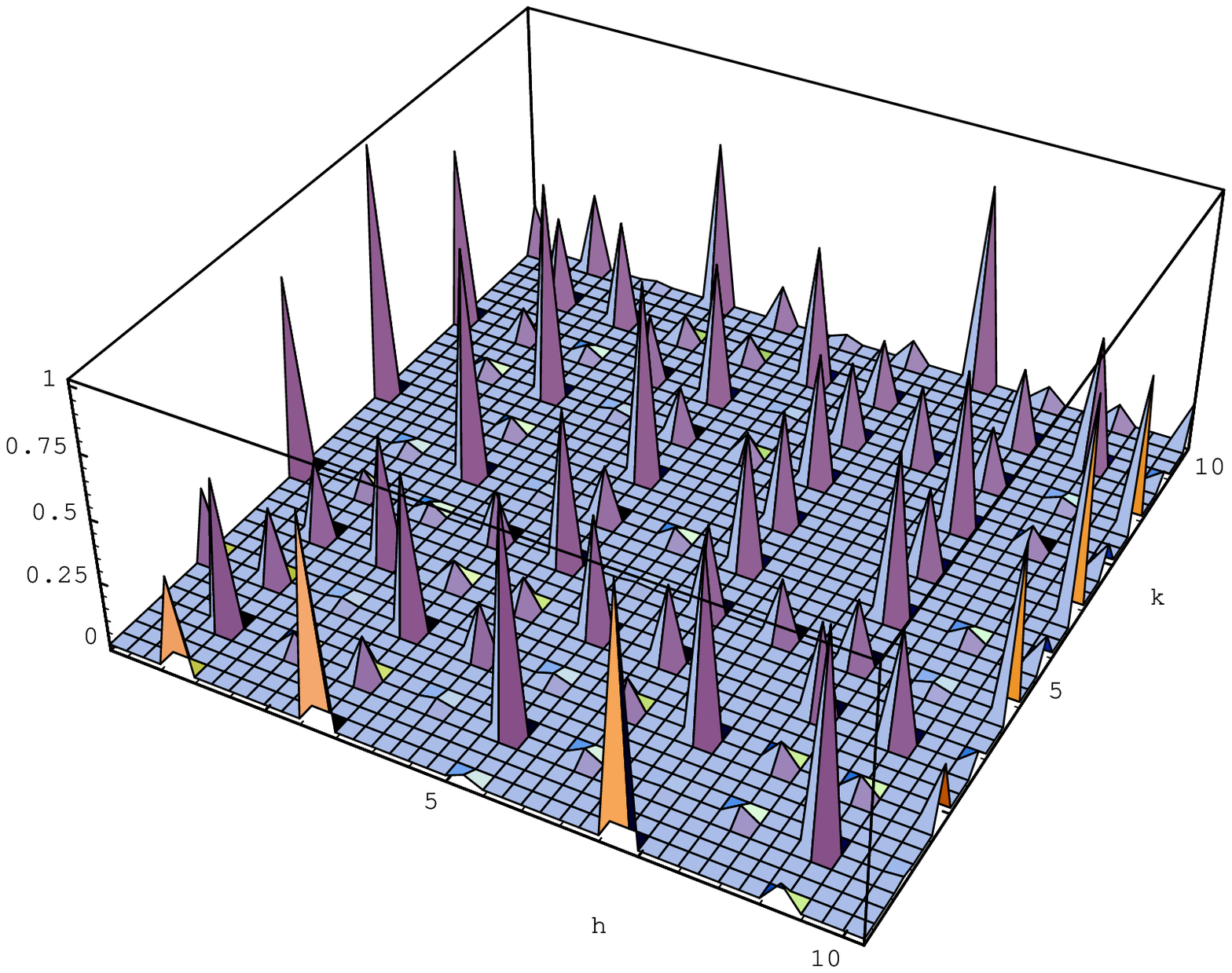,height=9.5cm,angle=0}}
\end{figure}

\centerline{\huge (b)} 
\vskip 0.5cm
\centerline{\huge Fig. 10, B. Normand and T. M. Rice }

\end{document}